\documentclass[twocolumn,english,superscriptaddress]{revtex4-1}
\usepackage[T1]{fontenc}
\usepackage[latin9]{inputenc}
\usepackage{amsmath}
\usepackage{amssymb}
\usepackage{graphicx}
\usepackage{babel}
\usepackage{lipsum}

\usepackage{ulem}

\usepackage{color}

\begin{document}
\title{Optomechanical Effects in Nanocavity-enhanced Resonant Raman Scattering of a Single Molecule}

\author{Xuan-Ming Shen}
\address{Henan Key Laboratory of Diamond Optoelectronic Materials and Devices, Key Laboratory of Material Physics, Ministry of Education, School of Physics and Microelectronics, Zhengzhou University, Daxue Road 75, Zhengzhou 450052, China}

\author{Yuan Zhang}
\email{yzhuaudipc@zzu.edu.cn}
\address{Henan Key Laboratory of Diamond Optoelectronic Materials and Devices, Key Laboratory of Material Physics, Ministry of Education, School of Physics and Microelectronics, Zhengzhou University, Daxue Road 75, Zhengzhou 450052, China}

\author{Shunping Zhang}
\affiliation{School of Physics and Technology, Center for Nanoscience and Nanotechnology, and Key Laboratory of Artificial Micro- and Nano-structures of Ministry of Education, Wuhan University, Wuhan 430072, China}

\author{Yao Zhang}
\affiliation{Hefei National Research Center for Physical Sciences at the Microscale and Synergetic Innovation Centre of Quantum Information and Quantum Physics, University of Science and Technology of China, Hefei, Anhui 230026, China}

\author{Qiu-Shi Meng}
\affiliation{Hefei National Research Center for Physical Sciences at the Microscale and Synergetic Innovation Centre of Quantum Information and Quantum Physics, University of Science and Technology of China, Hefei, Anhui 230026, China}

\author{Guangchao Zheng}
\affiliation{School of Physics and Microelectronics, Zhengzhou University, Daxue Road 75, Zhengzhou 450052}

\author{Siyuan Lv}
\address{Department of Physics, University of Science and Technology Beijing, 100083 Beijing, China}

\author{Luxia Wang}
\address{Department of Physics, University of Science and Technology Beijing, 100083 Beijing, China}

\author{Roberto A. Boto}
\address{Center for Material Physics (CSIC - UPV/EHU and DIPC) Paseo Manuel de Lardizabal 5, Donostia-San Sebastian Gipuzkoa 20018, Spain}
\address{Donostia International Physics Center, Paseo Manuel de Lardizabal 4, Donostia-San Sebastian 20018, Spain}

\author{Chongxin Shan}
\email{cxshan@zzu.edu.cn}
\address{Henan Key Laboratory of Diamond Optoelectronic Materials and Devices,
Key Laboratory of Material Physics, Ministry of Education, School of Physics and Microelectronics, Zhengzhou University, Daxue Road 75, Zhengzhou 450052, China}

\author{Javier Aizpurua }
\email{aizpurua@ehu.es}
\address{Center for Material Physics (CSIC - UPV/EHU and DIPC) Paseo Manuel de Lardizabal 5, Donostia-San Sebastian Gipuzkoa 20018, Spain}
\address{Donostia International Physics Center, Paseo Manuel de Lardizabal 4, Donostia-San Sebastian 20018, Spain}

\begin{abstract}
In this article, we  address the optomechanical effects in surface-enhanced resonant Raman scattering (SERRS) from a single molecule in a nano-particle on mirror (NPoM) nanocavity by developing a quantum master equation theory, which combines macroscopic quantum electrodynamics  and electron-vibration interaction within the framework of open quantum system theory. We supplement the theory with electromagnetic simulations and time-dependent density functional theory calculations in order to study the SERRS of a methylene blue molecule in a realistic NPoM nanocavity. The simulations allow us not only to identify the conditions to achieve conventional optomechanical effects, such as vibrational pumping, non-linear scaling of Stokes and anti-Stokes scattering, but also to discovery distinct behaviors, such as the saturation of exciton population, the emergence of  Mollow triplet side-bands, and higher-order Raman scattering. All in all, our study might guide further investigations of optomechanical effects in resonant Raman scattering. 
\end{abstract}
\maketitle

\section{Introduction}
Surface-enhanced Raman scattering (SERS) 
refers to the enhacement of the Raman signal of molecules located near metallic nanostructures~\citep{ECLeRu}. This effect is partially due to the charge transfer between the molecule and the metal (chemical enhancement ~\citep{JRLombardi}) but the main contribution is due to the enhanced electromagnetic field by the metallic nanostructure (electromagnetic field enhancement~\citep{MMoskovits}). Moreover, since Raman enhancement can reach tens of orders of magnitude for molecules near electromagnetic hot-spots~\citep{HXu}, vibrational pumping associated with the enhanced Stokes scattering can potentially compete with thermal vibrational excitation, and is thus able to introduce non-linear scaling of anti-Stokes scattering with increasing laser intensity~\citep{KKneipp,RCMaherJPCB,RCMaherCSR,MCGalloway}. 

Earlier studies on vibrational pumping were hindered by the difficulty of quantitatively estimating the vibrational pumping rate. To overcome this problem, in recent years,  M. K. Schmidt et al.~\citep{MKSchmidtACSNano} and P. Roelli, et al.~\citep{PRoelli} were inspired by cavity optomechanics~\citep{MAspelmeyer}, and developed a molecular optomechanics theory~\citep{MKSchmidtFaraday} for off-resonant Raman scattering. According to this theory, molecular vibrations interact with the plasmonic response of metallic nanostructures through optomechanical coupling, and the vibrational pumping rate can be quantitatively determined by  this coupling together with the molecular vibrational energy and the plasmonic response (such as mode energy, damping rate and laser excitation). Besides vibrational pumping, molecular optomechanics also allows us to investigate many novel effects, such as non-linear divergent Stokes scattering (known as parametric instability in cavity optomechanics~\citep{MAspelmeyer}), collective optomechanical effects~\citep{YZhangACSPhotonics}, higher-order Raman scattering~\citep{MKDezfouli}, and optical spring effect~\citep{WMDeacon} among others.

Metallic nanocavities, formed by metallic nanoparticle dimers~\citep{WZhu}, metal nano-particle on mirror (NPoM)~\citep{JJBaumberg} or STM tip-on-metallic substrate~\citep{XWang}, are ideal configuration for observation of novel optomechanical effects in experiments, because they provide hundred folds of local field enhancement inside the nano-gaps, and thus provide easily SERS enhancement over $10^{12}$. Using biphenyl-4-thiol molecules inside NPoM gold nanocavities, F. Benz, et al.~\citep{FBenzScience}  observed non-linear scaling of anti-Stokes SERS with increasing continuous-wave laser excitation (thus justifying vibrational pumping). Later on, using the same system, non-linear scaling of Stokes SERS was observed for pulsed laser illumination of much stronger intensity at room temperature~\citep{NLombardi} (proving the precursor of parametric instability and the collective optomechanical effect). Recently, Y. Xu, et al.~\citep{XuY} observed also the similar non-linear Stokes SERS with a MoS2 monolayer within metallic nanocube-on-mirror nanocavities.

Most of studies on molecular optomechanics so-far focus  on the off-resonant Raman scattering. Because off-resonant Raman scattering is usually weak, the observation of its optomechanical effects requires normally very strong laser excitation. To reduce the required laser intensity, in this article, we propose to combine the metallic nanocavities with the molecular resonant effect to enhance the Raman scattering of molecules. The molecular resonant effect refers to the enhancement of the Raman scattering when the laser is resonant with molecular electronic excited states. Indeed, the earliest studies on vibrational pumping focused on surface enhanced resonant Raman scattering (SERRS) of dye molecules~\citep{KKneipp,RCMaherJPCB,RCMaherCSR,MCGalloway} (such as crystal violet, or rhodamine 6G). 

In our previous works~\citep{TNeumanNP,TNeumanPRA}, we extended the molecular optomechanics approach based on single plasmon mode to SERRS, and showed that the electron-vibration coupling, leading to resonant Raman scattering, resembles the plasmon-vibration optomechanical coupling, and thus we show that optomechanical effects can also occur in resonant Raman scattering. Moreover, since the former coupling is usually much larger, and the electronic excitation is usually much narrower than the plasmon excitation, optomechanical effects in SERRS can potentially occur for much smaller laser intensities. Furthermore, in our other works~\citep{WMDeacon,YZhang}, we showed that the plasmonic response of metallic nanocavities is far more complex than that of a single mode, and the molecular optomechanics is strongly affected by the plasmonic pseudo-mode, formed by the overlapping higher-order plasmonic modes~\citep{ADelga}.

\begin{figure}[!htb]
\begin{centering}
\includegraphics[scale=0.5]{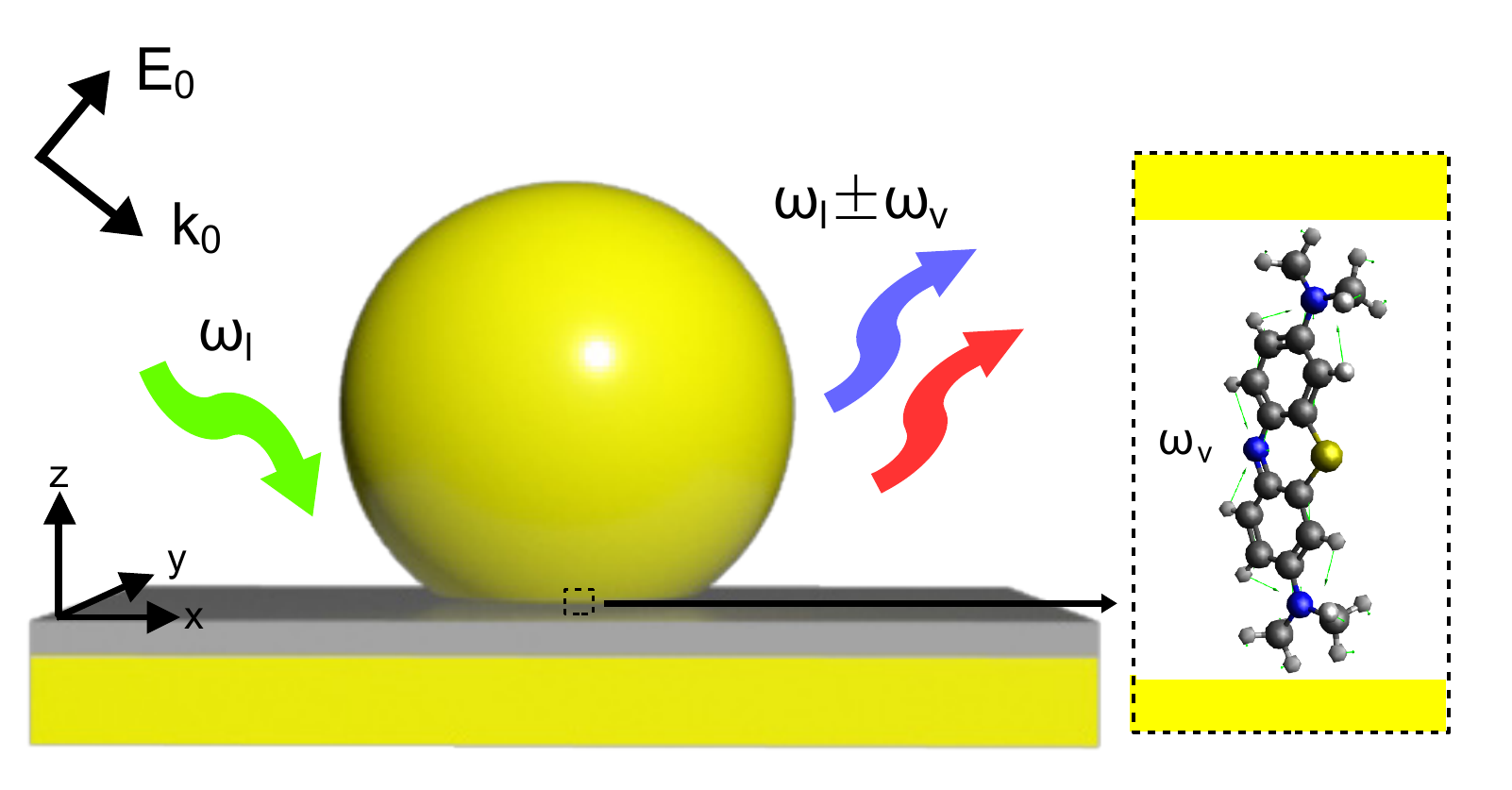}
\par\end{centering}
\caption{ \label{fig:NPoM} A vertically-orientated methylene blue molecule (with black, white, blue, yellow spheres for carbon, hydrogen, nitrogen, sulfur atoms, respectively) in the middle of a nano-particle on mirror (NPoM) nanocavity of $0.9$ nm thick, formed by a truncated gold sphere with $40$ nm diameter and bottom facet of $10$ nm diameter on top of a flat gold substrate. The laser excitation of frequency $\omega_l$  is enhanced in the nanocavity, and the enhanced local field excites the molecule vibrating with frequency $\omega_{\nu}$. The emitted field at frequency $\omega_l$  (Rayleigh scattering) and at frequencies $\omega_l - \omega_{\nu} $, $\omega_l + \omega_{\nu} $  (Stokes and anti-Stokes scattering), as well as frequencies independent of  $\omega_l$ (fluorescence), is enhanced and propagated to the far-field. }
\end{figure}

To address SERRS from molecules in realistic metallic nanocavities, here, we develop a theory that combines the macroscopic quantum electrodynamics description~ \citep{NRivera,SScheel} with the electron-vibration interaction, and derive a quantum master equation for the molecular electronic and vibrational dynamics. As an example, we apply our theory to a single methylene blue  molecule inside a gold NPoM nanocavity, as shown in Fig. \ref{fig:NPoM}. To maximize the methylene blue molecule-nanocavity interaction, we assume that the methylene blue molecule is encapsulated by a cucurbit[n] cage~\citep{RChikkaraddy} so that the molecule stands vertically. 

Our study shows that most of the optomechanical effects, such as vibrational pumping, parametric instability,  vibrational saturation, Raman line shift and narrowing and so on, can occur in SERRS at lower laser intensity threshold. However, the molecular excitation saturates for strong laser excitation because of its two-level fermionic nature (in contrast to the infinite-levels of a bosonic plasmon), and the SERRS signal saturates and even vanishes for strong laser excitation. In addition, we also find that the resonant fluorescence is red-shifted by about $40$ meV (plasmonic Lamb shift~\citep{YaoNat,BYang}), and broadened by about $22$ meV (due to the Purcell effect), and also shows three broad peaks for strong laser excitation~\citep{LyuS} (corresponding to the Mollow triplet similar to the situation in quantum optics~\citep{MOScully}). 

Our article is organized as follows. We present first our theory for  SERRS of single molecule in plasmonic nanocavities in Section \ref{sec:QME}, which is followed by the time-dependent density functional theory (TDDFT) calculation of the methylene blue molecule in Section \ref{sec:EVCoup} and the electromagnetic simulation of the NPoM nanocavity in Section \ref{sec:EMSim}. In Section \ref{sec:SERR}, we study the evolution of the SERRS and fluorescence with increasing laser illumination, which is blue-, zero- or red-detuned with respect to the molecular excitation, respectively. In the end, we conclude our work and comment on the extensions in future.

\section{Quantum Master Equation \label{sec:QME}}

To address the processes shown in Fig. \ref{fig:NPoM}, we have developed a theory that combines macroscopic quantum electrodynamics and electron-vibration interaction.
In Appendix \ref{sec:effmaseqn}, we detail the treatment of the interaction between a single molecule and the plasmonic (electromagnetic) field of  the metallic nanocavity. To reduce the degrees of freedom, we apply the open quantum system theory~\citep{HPBreuer} where we consider the plasmonic field as a reservoir and treat the molecule-plasmonic field interaction as a perturbation in second-order, to finally achieve an effective master equation for the single molecule. Here,  this treatment is valid since the single molecule couples relatively weakly with the nanocavity. However, further consideration is required for the system with more molecules, which might enter into the strong coupling regime~\citep{RChikkaraddy}.

To account for other mechanisms, like the molecular vibrations and the molecular excitation, we generalize the effective master equation to obtain
\begin{align}
\frac{\partial}{\partial t}\hat{\rho} & =-\frac{i}{\hbar}\left[\hat{H}_{ele}+\hat{H}_{las}+\hat{H}_{vib}+\hat{H}_{ele-vib}+\hat{H}_{pla},\hat{\rho}\right]\nonumber \\
 & + (\Gamma +\gamma_{e})\mathcal{D}_{\hat{\sigma}}\left[\hat{\rho}\right]+\frac{\chi_{e}}{2}\mathcal{D}_{\hat{\sigma}^{z}}\left[\hat{\rho}\right]\nonumber \\
 & +\sum_{\nu} \gamma_{\nu} \{\left(n_{\nu}^{th}+1\right)\mathcal{D}_{\hat{b}_{\nu}}\left[\hat{\rho}\right]+ n_{\nu}^{th}\mathcal{D}_{\hat{b}^{\dagger}_{\nu}}\left[\hat{\rho}\right]\}.\label{eq:me}
\end{align}
In the above equation, we treat the molecular electronic ground and excited state as a two-level system, and specify its Hamiltonian as $\hat{H}_{ele}=\hbar(\omega_{e}/2)\hat{\sigma}^{z}$ with  transition frequency $\omega_{e}$ and Pauli operator $\hat{\sigma}^{z}$. We treat the molecular excitation semi-classically with the Hamiltonian $\hat{H}_{\rm las}=\hbar\left(\hat{\sigma}^{\dagger}ve^{-i\omega_{las}t}+v^{*}e^{i\omega_{las}t}\hat{\sigma}\right)$, where the  molecule-near field coupling $\hbar v=-\mathbf{d}_{m}\cdot\mathbf{E}\left(\mathbf{r}_{m},\omega_{las}\right)$ is determined by the enhanced local electric field $\mathbf{E}\left(\mathbf{r}_{m},\omega_{las}\right)$ at the molecular position $\mathbf{r}_{m}$, activated by a laser with frequency $\omega_{las}$. Here, $\hat{\sigma}^{\dagger},\hat{\sigma}$ are the raising and lowering operator of the molecular excitation. We approximate the molecular vibrations as quantized harmonic oscillators and specify their Hamiltonian $\hat{H}_{vib}=\hbar \sum_{\nu} \omega_{\nu}\hat{b}^{\dagger}_{\nu} \hat{b}_{\nu}$ with the frequency $\omega_{\nu}$, the creation $\hat{b}^{\dagger}_{\nu}$ and annihilation $\hat{b}_{\nu}$ operators of the $\nu$-th vibrational mode. The electron-vibration interaction takes the form $\hat{H}_{ele-vib}=\hbar\sum_{\nu} \omega_{\nu} d_{\nu}\hat{\sigma}^{\dagger}\hat{\sigma}\left(\hat{b}^{\dagger}_{\nu}+\hat{b}_{\nu}\right)$ with the dimensionless displacement $d_{\nu}=\sqrt{S_{\nu}}$($S_{\nu}$ is known as the Huang-Rhys factor~\citep{KHuang}) of the parabolic potential energy surface for the electronic ground and excited state. Here, the electron transition frequency $\omega_e$ has already accounted for the shift $\sum_{\nu} d_{\nu} \omega_{\nu}$ due to the electron-vibration coupling.

The elimination of the plasmonic field leads to one Hamiltonian $\hat{H}_{pla}= \hbar(\Omega/2)\hat{\sigma}^{z}$ and one Lindblad term $\Gamma\mathcal{D}_{\hat{\sigma}} \left[\hat{\rho}\right]$ with the superoperator $\mathcal{D}_{\hat{o}} \left[\hat{\rho}\right] =\hat{o}\hat{\rho} \hat{o}^{\dagger}-\frac{1}{2}\hat{o}^{\dagger}\hat{o}\hat{\rho}-\frac{1}{2}\hat{\rho} \hat{o}^{\dagger}\hat{o}$ (for any operator $\hat{o}$), which describe the shift 
\begin{equation}
\Omega=-\frac{1}{\hbar\epsilon_{0}}\frac{\omega_{e}^{2}}{c^{2}}\mathbf{d}_{m}\cdot\mathrm{Re}\overleftrightarrow{G}\left(\mathbf{r}_{m},\mathbf{r}_{m};\omega_{e}\right)\cdot\mathbf{d}_{m}^{*} \label{eq:Omega}
\end{equation}of the transition frequency (plasmonic Lamb shift~\citep{YaoNat,BYang}),  and the Purcell-enhanced decay rate
\begin{equation}
\Gamma=\frac{2}{\hbar\epsilon_{0}}\frac{\omega_{e}^{2}}{c^{2}}\mathbf{d}_{m}\cdot\mathrm{Im}\overleftrightarrow{G}\left(\mathbf{r}_{m},\mathbf{r}_{m};\omega_{e}\right)\cdot\mathbf{d}_{m}^{*} \label{eq:Gamma}
\end{equation}of the molecule. Here, $\epsilon_{0},c$ are the vacuum permittivity and the speed of light, and $\mathbf{d}_{m}$ is the molecular transition dipole. $\overleftrightarrow{G}\left(\mathbf{r},\mathbf{r}';\omega\right)$ is the  dyadic Green's function, which  connects usually  the electric field with frequency $\omega$  at the position $\mathbf{r}$ with the dipole point source at another position $\mathbf{r}'$ in classical electrodynamics. The remaining terms of Eq. \eqref{eq:me} describe the dissipation of the electronic states, including the non-radiative decay $\gamma_{e}\mathcal{D}_{\hat{\sigma}}\left[\hat{\rho}\right]$ with rate $\gamma_{e}$, and the dephasing $\frac{\chi_{e}}{2}\mathcal{D}_{\hat{\sigma}^{z}}\left[\hat{\rho}\right]$ with rate $\chi_{e}$, and the thermal decay and pumping of the vibrational modes  $\sum_{\nu} \gamma_{\nu}\left(n_{\nu}^{th}+1\right)\mathcal{D}_{\hat{b}}\left[\hat{\rho}\right]+\gamma_{\nu}n_{\nu}^{th}\mathcal{D}_{\hat{b}^{\dagger}}\left[\hat{\rho}\right]$ with rate $\gamma_{\nu}$, where $n_{\nu}^{th}=\left[{\rm exp}\left\{ \hbar\omega_{\nu}/k_{B}T\right\} -1\right]^{-1}$ is the thermal vibrational population at temperature $T$ ($k_{B}$ is Boltzmann constant).  

In Appendix \ref{sec:spe}, we have derived the formula to compute the spectrum $dW(\omega)/d\Omega$  measured by a detector in the far-field:
\begin{equation}
\frac{dW}{d\Omega}\left(\omega\right)\approx K(\omega) \mathrm{Re}\int_{0}^{\infty}d\tau e^{i\omega\tau}\mathrm{tr}\left\{ \hat{\sigma}\hat{\varrho}\left(\tau\right)\right\}.  \label{eq:SpeFormula}
\end{equation}
In this equation, the propagation factor is defined as 
\begin{equation}
K(\omega)=\frac{\hbar^{2}cr^{2}}{4\pi^{2}\epsilon_{0}}\left|\frac{\omega^{2}}{c^{2}}\overleftrightarrow{G}\left(\mathbf{r}_d,\mathbf{r}_{m};\omega\right)\cdot\mathbf{d}^*_{m}\right|^{2},
\end{equation}where the dyadic Green's function  $\overleftrightarrow{G}\left(\mathbf{r}_d,\mathbf{r}_{m};\omega\right)$ connects the electric field at frequency $\omega$ at position ${\bf r}_d$ of the detector with the molecule at the position $\mathbf{r}_{m}$. $r$ denotes the distance between the molecule and the detector. In addition, the operator $\hat{\varrho}\left(\tau\right)$ satisfies the same master equation as $\hat{\rho}$, but with initial condition $\hat{\varrho}\left(0\right)=\hat{\rho}\hat{\sigma}^{\dagger}$. 

To solve the master equation [Eq.(\ref{eq:me})], we introduce the density matrix $\rho_{an,bm}$ (with $a,b=g,e$) in the basis of the product states $\left\{ \left|e\right\rangle \left|m\right\rangle ,\left|g\right\rangle \left|n\right\rangle \right\} $, where $\left|e\right\rangle ,\left|g\right\rangle $ denote the electronic
excited and ground state of the molecule, and $\left|m\right\rangle =\prod_\nu \left|m_\nu\right\rangle  ,\left|n\right\rangle =\prod_\nu \left|n_\nu\right\rangle $  are the occupation number states of the vibrational modes ($m_\nu,n_\nu$ are positive integers). From Eq. (\ref{eq:me}), we can easily derive the equation for the density matrix. In the current study, we utilize the QuTip toolkit~\citep{JRJohansson2012,JRJohansson2013} to solve the density matrix equation. By solving this equation, we can determine the electronic excited state population $\left \langle \hat{\sigma}^\dagger \hat{\sigma} \right \rangle ={\rm tr}\{\hat{\sigma}^\dagger \hat{\sigma} \hat{\rho}\}$, the mean vibrational population $ \left \langle \hat{b}_{\nu}^\dagger \hat{b}_{\nu} \right \rangle= {\rm tr}\{ \hat{b}_{\nu}^\dagger \hat{b}_{\nu} \hat{\rho}\}$, the two-time correlation $C(\tau) \equiv {\rm tr}\left\{ \hat{\sigma}\hat{\varrho}\left(\tau\right)\right\}$, and the spectrum $dW(\omega)/d\Omega$.    

In the weak-excitation limit $\left \langle \hat{\sigma}^\dagger \hat{\sigma} \right \rangle\ll 1$, we can apply the Holstein-Primakoff approximation~\citep{THolstein} to the molecule by introducing the bosonic creation $\hat{a}$ and annihilation operator $\hat{a}^\dagger$ such that: $\hat{\sigma}^z\equiv 2\hat{a}^\dagger \hat{a}  -1 $, $\hat{\sigma}^\dagger\equiv \hat{a}^\dagger \sqrt{1-\hat{a}^\dagger \hat{a}}\approx \hat{a}^\dagger$,$\hat{\sigma} \equiv \sqrt{1-\hat{a}^\dagger \hat{a}} \hat{a} \approx \hat{a}$. In this case, we can carry out the replacement $\hat{\sigma}^z \to \hat{a}^\dagger \hat{a}$, $\hat{\sigma}^\dagger \to \hat{a}^\dagger$, $\hat{\sigma} \to \hat{a}$ in the master equation given above, and then the resulted equation is equivalent to the one used in the molecular optomechanics theory~\citep{MKSchmidtACSNano,PRoelli}. This indicates that the molecular excitation works effectively as the plasmon mode, and the electron-vibration coupling as the optomechanical coupling i.e. $g_{\nu} = \omega_{\nu} d_{\nu}$. Thus, we expect to find many interesting optomechanical effects in the current system, namely vibrational pumping, non-linear Raman scattering, Raman line-shift and broadening~\citep{YZhangACSPhotonics} and so on. For typical value of $\hbar \omega_{\nu}=200$ meV and $d=0.1$, we obtain $\hbar g_{\nu} =20$ meV, which is about two orders of magnitude larger than in typical molecular optomechanics~\citep{MKSchmidtACSNano,YZhangACSPhotonics}. Thus, we  expect to observe a variety of optomechanical effects in SERRS with however lower laser intensity threshold.

\begin{figure}[t]
\begin{centering}
\includegraphics[scale=0.5]{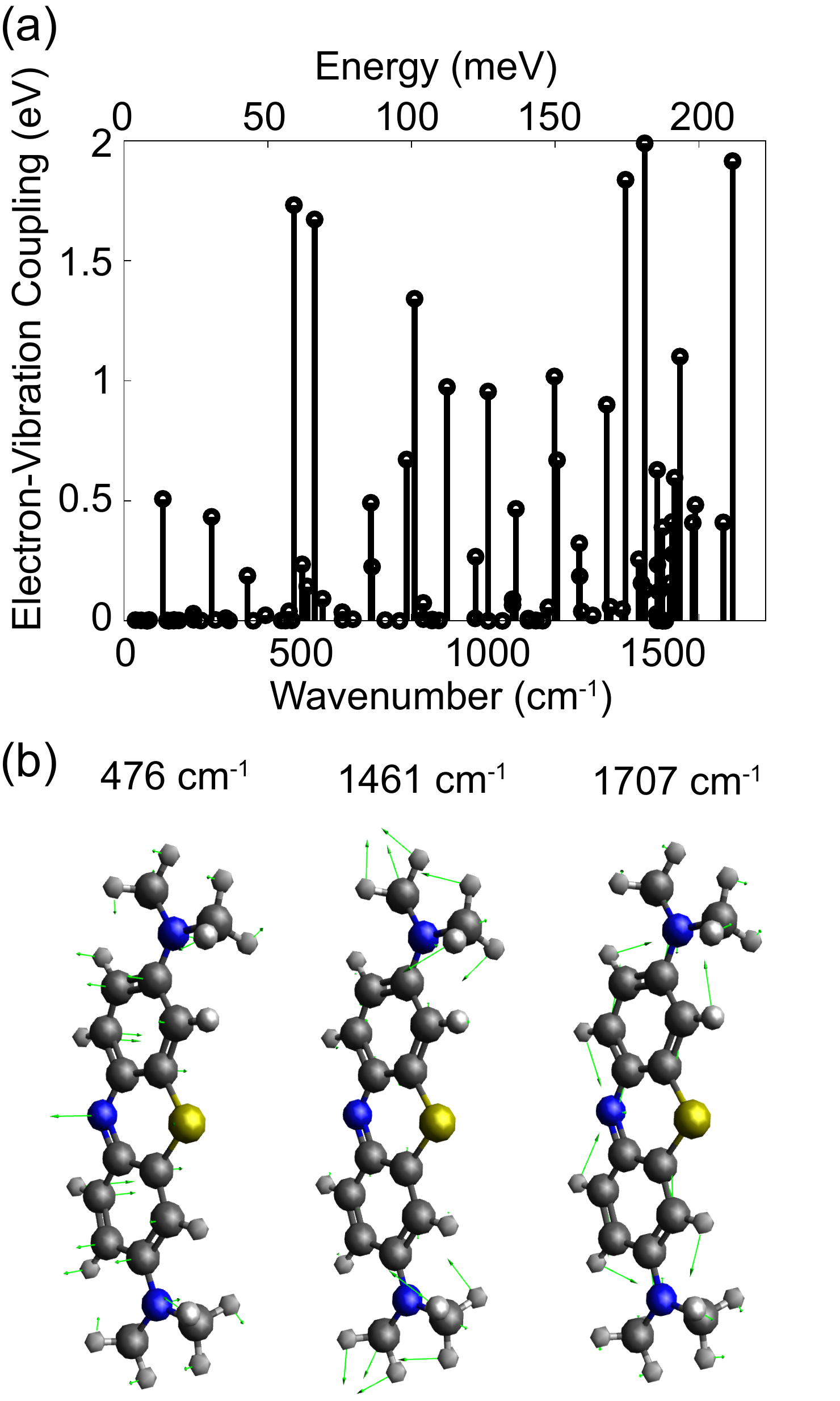}
\par\end{centering}
\caption{ \label{fig:DFT} Electronic and vibrational calculation of the methylene blue molecule. (a) Electron-vibration coupling versus wavenumber (lower axis), energy (upper axis) of the vibrational modes. (b) Vibrational patterns of the $476$ ${\rm cm}^{-1}$(left), $1461$ ${\rm cm}^{-1}$ (middle) and $1707$ ${\rm cm}^{-1}$ (right) vibrational modes of the same molecule. 
}
\end{figure}

\section{Electronic and Vibrational Analysis of Methylene Blue molecule \label{sec:EVCoup}}

To apply the theory developed in the previous section, we carry out electronic and  vibrational excitation analysis for the methylene blue molecule with the use of the Gaussian 16 package, and utilize the cam-B3LYP hybrid functional and 6-311g(d,p) basis set (Fig. \ref{fig:DFT}). Firstly, we carry out Density Functional Theory (DFT) calculations to optimize the molecule, compute the vibrational modes for the electronic ground state, and extract their equilibrium normal-mode coordinates $Q_{\nu}$. Secondly, we carry out time-dependent DFT calculations to optimize the molecule on the first excited state, compute the vibrational modes for this state, and extract the corresponding equilibrium normal-mode coordinates $Q_{\nu}'$. From these calculations, we can also determine the transition energy $\hbar\omega_e$, and the transition dipole moment $\mathbf{d}_{m}$ for the maximal emission (in the Tamm-Dancoff approximation \citep{SHirata}). Finally, we compute the vibronic spectrum for one-photon emission using the Franck-Condon approach\citep{YHu}, and extract the displacement matrix $K$ in the Duschinsky transformation\citep{FDuschinsky} $Q'_{\nu}=JQ+K$ (with $J$ matrix representing the normal mode mixing during the transition). The Huang-Rhys factors can be computed according to the relationship $S_{\nu} = (4\pi^2 \omega_{\nu} c/2h) K_{\nu}^2$.

From the above calculations, we obtain the transition energy $\hbar\omega_e\approx 2.88$ eV, which is consistent with the result in~\citep{TNeuman2018} but higher than the energy of the emission maximum $1.82$ eV (wavelength of $680$ nm) due to the limitation of TDDFT to describe $\pi$-conjugated systems~\citep{TBDQueiroz}. The transition dipole moment obtained is $|\mathbf{d}_{m}|\approx 11$ Debye (along the long molecular axis) for the methylene blue molecule, which agrees with the result in Ref. [\citenum{TNeuman2018}] but is larger than the values $2.55 \sim 4.23$ Debye in Ref. [\citenum{TBDQueiroz}]. Fig. \ref{fig:DFT}(a) shows the computed electron-vibration coupling in the order of ${\rm eV}$ versus the wavenumber of vibrational modes. This coupling  is strong for the vibrational modes around $500$ ${\rm cm}^{-1}$, $1400$ ${\rm cm}^{-1}$  and  $1700$ ${\rm cm}^{-1}$, and this characteristic is similar to that of the observed SERRS of the same molecule in experiments (see Fig. S11 in Ref.\citep{RChikkaraddy}). Fig. \ref{fig:DFT}(b) shows the vibrational pattern for the $476$ ${\rm cm}^{-1}$ (left), $1461$ ${\rm cm}^{-1}$ (middle) and $1707$ ${\rm cm}^{-1}$ (right) mode, which can be assigned to the ${\rm N-C}$ swing mode, the ${\rm CH_3}$-rocking mode, and the ${\rm C=C}$ stretching mode, respectively.  The electron-vibration coupling for these modes  is estimated as $\hbar \omega_{\nu} d_{\nu} = 1.732$ eV, $1.989$ eV, and  $1.915$ eV  from Fig. \ref{fig:DFT}(a).   

In the following simulations, we choose electronic transition energy $\hbar \omega_e = 1.82$ eV as measured in the experiment, and a transition dipole moment $|\mathbf{d}_{m}|\approx 3.8$ Debye (compromised value among different calculations), and downscale the electron-vibration coupling to the typical values, $\hbar \omega_{\nu} d_{\nu} = 17.3,19.9, 19.2$ meV for the $476  {\rm cm}^{-1}$, $1461 {\rm cm}^{-1}$ and $1707 {\rm cm}^{-1}$ mode (scaled down by a factor $100$  to reach the typical values), respectively. 

\begin{figure}[t]
\begin{centering}
\includegraphics[scale=0.45]{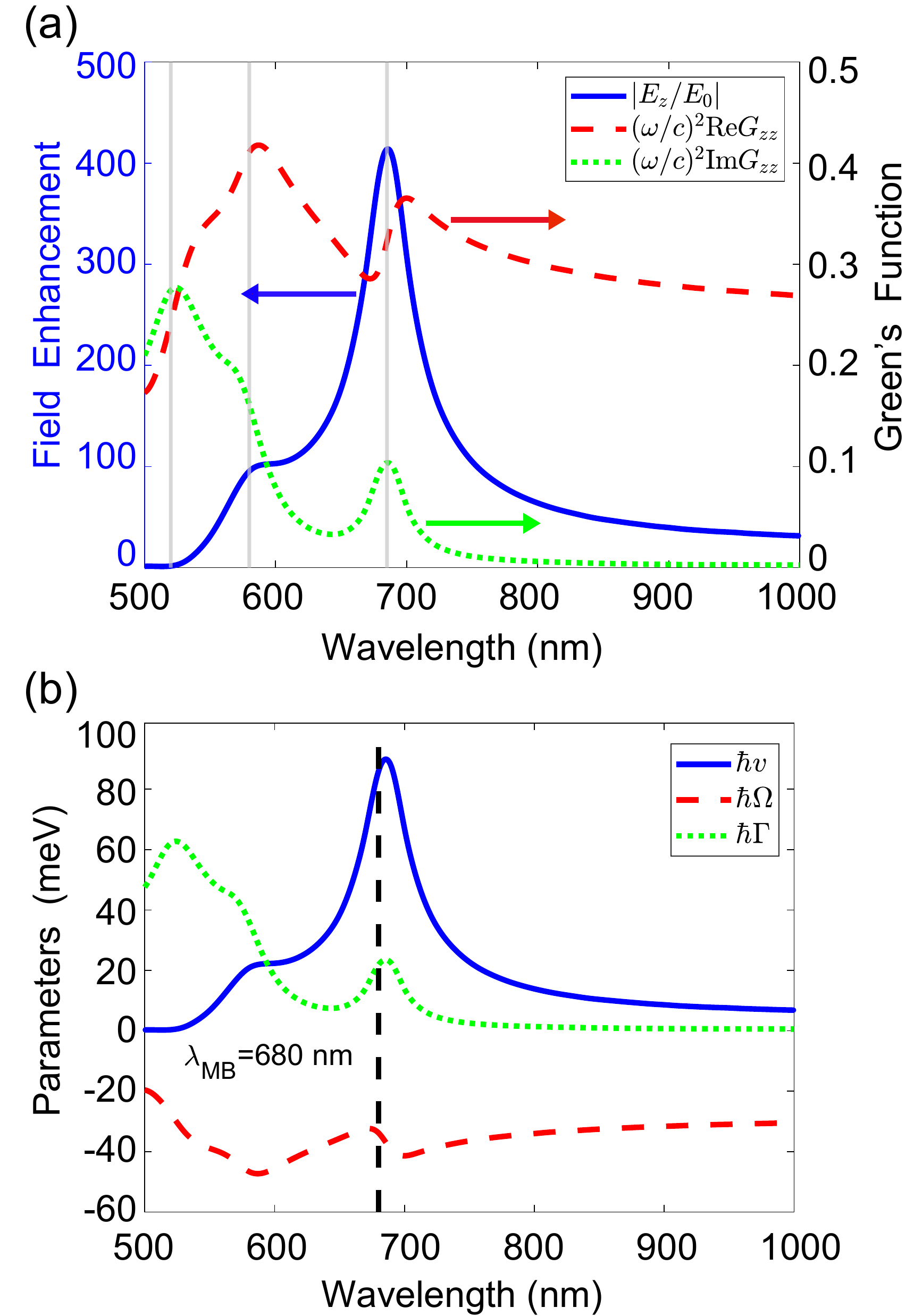}
\par\end{centering}
\caption{ \label{fig:EMSim} Plasmonic response of the NPoM nanocavity. (a) Near-field enhancement (blue solid line, left axis), real part (red dashed line) and imaginary part (green dotted line) of the scattered dyadic Green's function (right axis) for a dipole in the middle of the nanocavity, where the vertical gray lines from right to left show the wavelengths of $685$ nm, $580$ nm and $520$ nm, respectively. (b) Molecule-near field coupling $\hbar |v|$ (blue solid line) for laser intensity $10^4 \mu W/ \mu m^2$ versus the laser wavelength, Lamb shift $\hbar \Omega$ (red dashed line), and Purcell-enhanced decay rate $\hbar \Gamma$ (green dotted line)  versus the wavelength. Here, the molecule is assumed to stand vertically and the transition dipole is $3.8$ D, and the vertical dashed line indicates the  transition wavelength $\lambda_{e}= 680$ nm of the methylene blue molecule. For more details, see the text.}
\end{figure}

\section{Electromagnetic Response of the NPoM Nanocavity \label{sec:EMSim}}

The electromagnetic response of the NPoM nanocavity specified in Fig. \ref{fig:NPoM}, is shown in Fig. \ref{fig:EMSim}. First, we illuminate the system by a p-polarized plane-wave within a wavelength range  $[500,1000]$ nm with an incident angle of $55^{\rm o}$  to the surface normal, and compute the near-field enhancement of the vertical field component $|E_z/E_0|$ in the middle of the nanocavity [blue solid line in Fig. \ref{fig:EMSim}(a)]. The field enhancement shows two peaks with maxima about $400$ and $100$ at around $685$ nm, and $580$ nm, which are attributed to the bonding dipole plasmon (BDP) and the bounding quadrupole plasmon (BQP), respectively~\citep{FBenzScience, NLombardi, YZhang}.  In addition, we observe also two peaks at similar wavelengths in the far-field scattering spectrum (not shown). 

Second, we illuminate the system with a point-dipole located in the middle of the NPoM nanocavity, then compute the scattered electric field at the position of the dipole, and finally compute the scattered dyadic Green's function as the ratio of the scattered field and the dipole amplitude~\citep{ YZhang}. On the right axis of Fig. \ref{fig:EMSim}(a), we show  the real (red dashed line) and imaginary (green dotted line) part of the zz-component of the dyadic Green's tensor in such situation. The imaginary part shows two peaks at similar wavelengths as for the near-field enhancement, but also one extra peak at around $520$ nm, which can be attributed to the plasmon pseudomode arising from the overlapped higher-order plasmon modes~~\citep{ADelga}. In that figure, we also see two Fano-features around the wavelengths of the BDP and BQP mode in the real part of the dyadic Green's function. Note that the real part is always positive. As shown in our previous studies~\citep{YZhang}, the general wavelength-dependence of the dyadic Green's function resembles that of the metal-insulator-metal structure (for the same metal, same dielectric, and same thickness as the NPoM gap), which can be attributed to the resemblance of the NPoM mode with the guiding surface plasmon modes and the high-momentum surface-wave modes~\citep{GWFord}.

Assuming that the molecular transition dipole is $3.8$ D (typical value for the methylene blue molecule), we have computed and shown in Fig. \ref{fig:EMSim}(b) the molecule-near field coupling $\hbar v$ for laser intensity  $I_{las} = 10^4 \mu W/\mu m^2$ (blue solid line), a value achievable in experiment~\citep{NLombardi}, the plasmonic Lamb shift $\hbar \Omega$ (red dashed line) and the Purcell-enhanced spontaneous emission rate $\Gamma$ (green dotted line). $\hbar v$ follows the shape of the near-field enhancement, and reaches the maximal value around $80$ meV for the laser resonant with the BDP mode, i.e.  $\lambda_{\rm las} \approx \lambda_{\rm BDP}$. $\hbar \Omega$ follows the shape of the real part of the dyadic Green's function (with a sign change), and changes in the range of $[-20 $ ${\rm meV},-50 $ ${\rm meV}]$. $\hbar \Gamma$ follows the shape of the imaginary part of the dyadic Green's function, and varies in the range of $[0$, $60 $ ${\rm meV}]$. For the methylene blue molecule with the transition wavelength $\lambda_{e} = 680$ nm, we obtain $\hbar v\approx 80$ meV, $\hbar \Omega \approx -40$ meV and $\hbar \Gamma\approx 22$ meV. Since these parameters (and also the molecular intrinsic decay rate) are of the same order of magnitude, we expect that the molecule in the NPoM nanocavity can reach the non-linear regime with the typical laser intensities used in experiments. 

\section{SERRS Response to Laser Excitation \label{sec:SERR}}
We are now in the position to study the SERRS response of the methylene blue molecule in the NPoM nanocavity to the laser excitation. As explained in the end of Sec. \ref{sec:QME}, the theory for SERRS resembles that of molecular optomechanics in the low excitation limit, and thus we expect a different response for laser excitation which is resonant, blue- and red-detuned with respect to the molecular excitation, similar to the situations in molecular optomechanics~\citep{YZhangACSPhotonics}. Thus, in the following, we analyze the SERRS response in the three cases.

\begin{figure}
\begin{centering}
\includegraphics[scale=0.50]{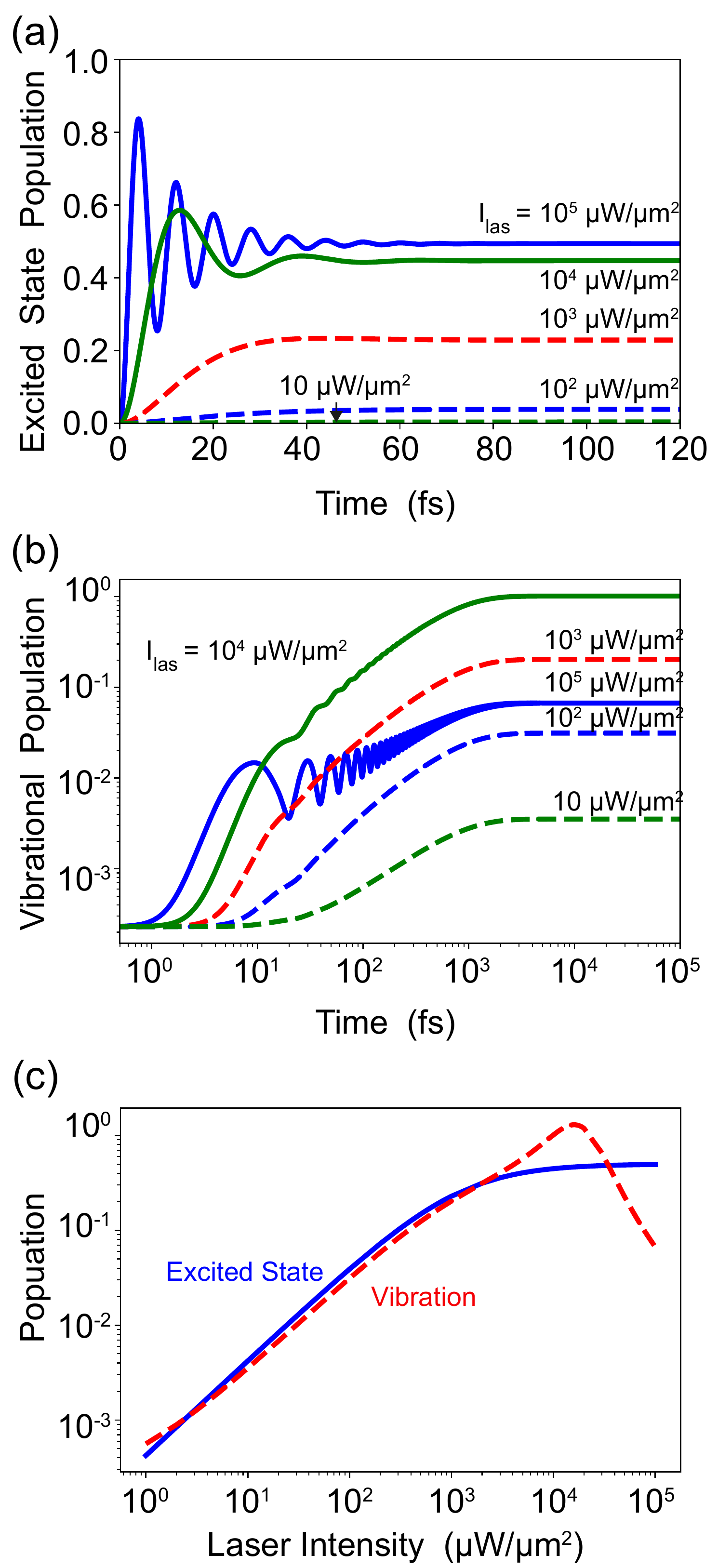}
\par\end{centering}
\caption{ \label{fig:fig4} Electronic and vibrational response to $\lambda_{\rm las}=693$ nm laser illumination with increasing intensity $I_{las}$, which is resonant to the Lamb-shifted molecular excitation and switched on abruptly at time zero. (a,b) Dynamics of the electronic excited-state population $\left \langle \hat{\sigma}^\dagger \hat{\sigma}\right \rangle$ (a) and vibrational population $\left \langle \hat{b}_{\nu}^\dagger \hat{b}_{\nu} \right \rangle$ (b). (c) Evolution of the quantities in (a) and (b) at the steady-state with increasing $I_{las}$.  Here, we consider the $1707 {\rm cm}^{-1}$ vibrational mode (with energy $\hbar\omega_{\nu}\approx211.6$ meV), which leads to a thermal population $n_{\nu}^{th}=2.29\times10^{-4}$ at room temperature $T=293$, and assume the electron-vibration coupling $\hbar\omega_{\nu}d_{\nu} \approx 19.2$ meV, the intrinsic excitonic decay rate $\hbar\gamma_e=56.89$ meV, the vanishing dephasing rate $\hbar\chi_e =0$, and the intrinsic vibrational decay rate $\hbar \gamma_{\nu}=1$ meV. For more information, see the text.}
\end{figure}

\subsection{Resonant Laser Excitation}

We study the SERRS response for continuous laser illumination of increasing intensity $I_{las}$,  which is resonant  with the molecular excitation. Figure. \ref{fig:fig4}(a) shows that for $I_{las}$  smaller than $10^3 \mu W/ \mu m^2$, the excited state population $\left \langle \hat{\sigma}^\dagger \hat{\sigma} \right \rangle$ increases firstly monotonously from zero, and then saturates at some finite values (dashed lines). For much larger $I_{las}$, the population shows oscillatory behavior before reaching the saturation value, and the Rabi oscillations become faster with increasing $I_{las}$. In addition, the saturated value also increases with increasing $I_{las}$, and eventually becomes close to $0.5$. These results indicate that for sufficiently large $I_{las}$, the coherent excitation of the molecule overcomes the intrinsic and Purcell-enhanced decay rates, and the molecule is driven to a superposition state of the electronic ground and excited state during the oscillatory period, reaching eventually to a mixed steady-state.

Figure \ref{fig:fig4}(b) shows that for $I_{las} \le 10^3 \mu W/\mu m^2$, the vibrational population $\left \langle \hat{b}_{\nu}^\dagger \hat{b}_{\nu} \right \rangle$ increases monotonously with time from the thermal value $n_{\nu}^{th}=2.29\times10^{-4}$ (at room temperature), and finally saturates at some finite value. The starting time for the rising of the vibrational population decreases with increasing $I_{las}$, and the saturation value increases. For much larger $I_{las}$,  the vibrational population shows some oscillations before reaching the saturation values. As $I_{las}$ increases, the oscillation becomes more pronounced, and the final saturated value decreases. The time to reach the saturated vibrational population is similar for all the laser intensities. Since the saturated values are much larger than the thermal value $n_{\nu}^{th}\approx 2.26\times 10^{-4}$, the laser illumination can pump significantly the molecular vibration via the electronic excitation.

The above results demonstrate the transient dynamics of the excited state and vibrational population. In the following, we summarize the influence of the laser intensity on these populations for the system at the steady-state [Fig. \ref{fig:fig4}(c)]. We can observe that as $I_{las}$ increases the electronic excited-state population (blue solid line) increases firstly linearly from zero for $I_{las}\le 10^2 \mu W/\mu m^2$, then sub-linearly for $I_{las}\le 10^4 \mu W/\mu m^2$ and finally saturates at value around $0.5$ for $I_{las}\le 10^5 \mu W/\mu m^2$. In contrast, the vibrational population (red dashed line) increases firstly sub-linearly for $I_{las}\le 10 \mu W/\mu m^2$ (due to the competition with the thermal excitation), linearly for $I_{las}\le 10^3 \mu W/\mu m^2$ (due to the vibrational pumping), then sub-linearly again for $I_{las}\le 10^4 \mu W/\mu m^2$, and finally it gets dramatically decreased for $I_{las}\le 10^5 \mu W/\mu m^2$ (due to the saturation of the molecular excitation).

\begin{figure}
\begin{centering}
\includegraphics[scale=0.53]{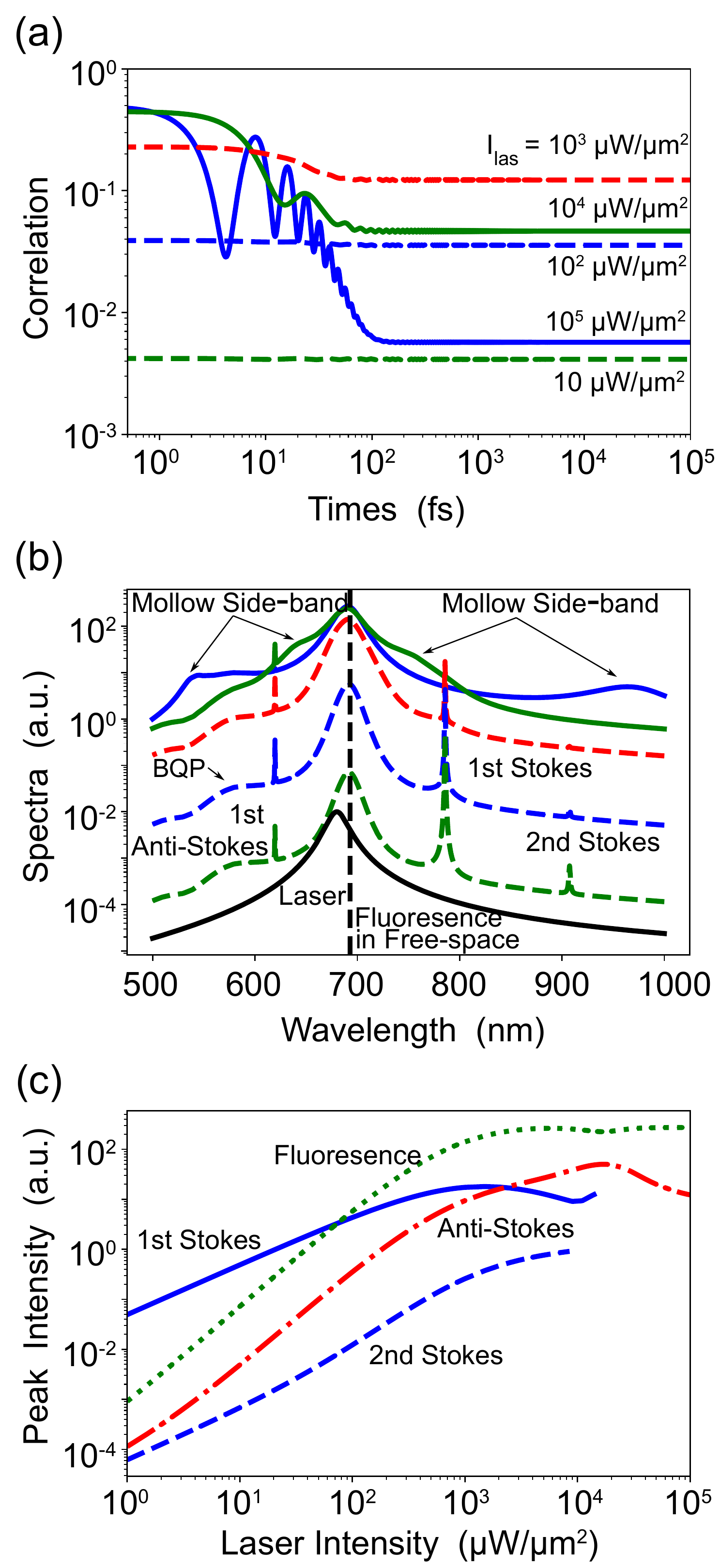}
\par\end{centering}
\caption{ \label{fig:fig5} SERRS response to $\lambda_{\rm las}=693$ nm laser illumination with increasing intensity $I_{las}$, resonant to the Lamb-shifted molecular excitation, which is switched on abruptly at time zero. (a) Dynamics of the two-time correlation function $C(\tau)$, and (b) SERRS spectra for $I_{las}=10,10^2,10^3 ,10^4,10^5 \mu W/\mu m^2$ [lower to upper curves following the same color code as in (a) ], together with molecular fluorescence in free-space (black solid line). (c) Evolution of the 2nd and 1st order Stokes line (blue dashed and solid line), and the anti-Stokes line (red dash-dotted line), as well as the fluorescence peak (green dotted line) with increasing $I_{las}$. In the panel (b) the Rayleigh scattering line at $\lambda_{\rm las}$ is removed. Other parameters are the same as in Fig. \ref{fig:fig4}. }
\end{figure}

After understanding the evolution of population, we now consider the response of the SERRS spectrum, as shown in Fig. \ref{fig:fig5}.  Figure \ref{fig:fig5}(a) shows that the two-time correlation $C(\tau)$ behaves similarly as the excited state population (as expected from the quantum regression theorem) except that it starts initially from the excited state population at the steady-state, i.e. $C(0) = {\rm tr}\left\{\hat{\sigma}^{\dagger} \hat{\sigma}\hat{\rho}_{ss} \right\}=\left \langle \hat{\sigma}^\dagger \hat{\sigma} \right \rangle$ (with the steady-state density operator $\hat{\rho}_{ss}$). Applying the Fourier transform to the correlation and weighting the results with the propagation factor according to Eq. \eqref{eq:SpeFormula}, we obtain the spectra shown in Fig. \ref{fig:fig5}(b). Here, we have removed the sharp line at the laser wavelength $\lambda_{\rm las}$ due to the elastic Rayleigh scattering. All the spectra show a broad fluorescence peak at around $690$ nm, which is about $12$ nm red-shifted from the molecular fluorescence peak in free-space (black solid line), due to the plasmonic Lamb shift. For larger laser intensity $I_{las} \ge 10^3 \mu W/\mu m^2$, this peak becomes much broader, and the fluorescence also shows two side peaks at shorter and longer wavelengths. We attribute these three peaks to the typical structure of a Mollow triplet~\citep{MOScully}, and the two side peaks to the molecule-laser (or plasmon) dressed states. For laser intensity $I_{las} \le 10^4 \mu W/\mu m^2$, we observe also three sharp lines at around $619$ nm, $785$ nm, and $907$ nm, which are due to the anti-Stokes, 1st order Stokes, and 2nd order Stokes scattering, respectively. The 1st order Stokes line is much stronger than the fluorescence peak and the  anti-Stokes line for the smallest $I_{las}$. However, with increasing  $I_{las}$, this line is overtaken firstly by the fluorescence peak, and then by the anti-Stokes line. In addition, the 2nd order Stokes line gets smaller and eventually vanishes with respect to the broad fluoresence for increasing $I_{las}$. Notice that the peak around $580$ nm is due to the radiation of the BQP mode.

Furthermore, in Fig. \ref{fig:fig5}(c), we investigate the fluorescence maximum (the green dotted line) and the Raman lines (other lines) as a function of  laser intensity $I_{las}$. We see that the fluorescence maximum follows the trend of the excited state population described in Fig.\ref{fig:fig4}(c), and the anti-Stokes line (red dash-dotted line) follows that of the vibrational population. Both show super-linear scaling for $I_{las}\le 10^3 \mu W/\mu m^2$. In contrast, the 1st order and 2nd order Stokes lines (blue solid and dashed-line) show linear scaling for $I_{las}\le 10^2 \mu W/\mu m^2$. For much larger $I_{las}$, the former shows sub-linear scaling, saturation and finally reduction, while the latter shows the evolution from super-linear to linear, and then to sub-linear. In addition, two critical laser intensities appear to be relevant in the evolution of the spectral peaks:  the 1st order Raman line is equally strong as the fluorescence  at about $10^2 \mu W/ \mu m^2$, and  as the anti-Stokes line at $2\times 10^3 \mu W/ \mu m^2$. Interestingly, for $I_{las}\ge 10^4 \mu W/\mu m^2$, we can still observe the anti-Stokes line but not the Stokes line. 

In Fig. \ref{fig:RS-res}(a,b) of Appendix \ref{sec:SupRes}, we have further examined the evolution of the Raman lineshape, and found that the Raman shift does not change while the Raman linewidth gets broader for larger $I_{las}$. In addition, the 2nd order Stokes line is about twice broader as a compared to the 1st order Stokes line. 

\begin{figure}[!htb]
\begin{centering}
\includegraphics[scale=0.53]{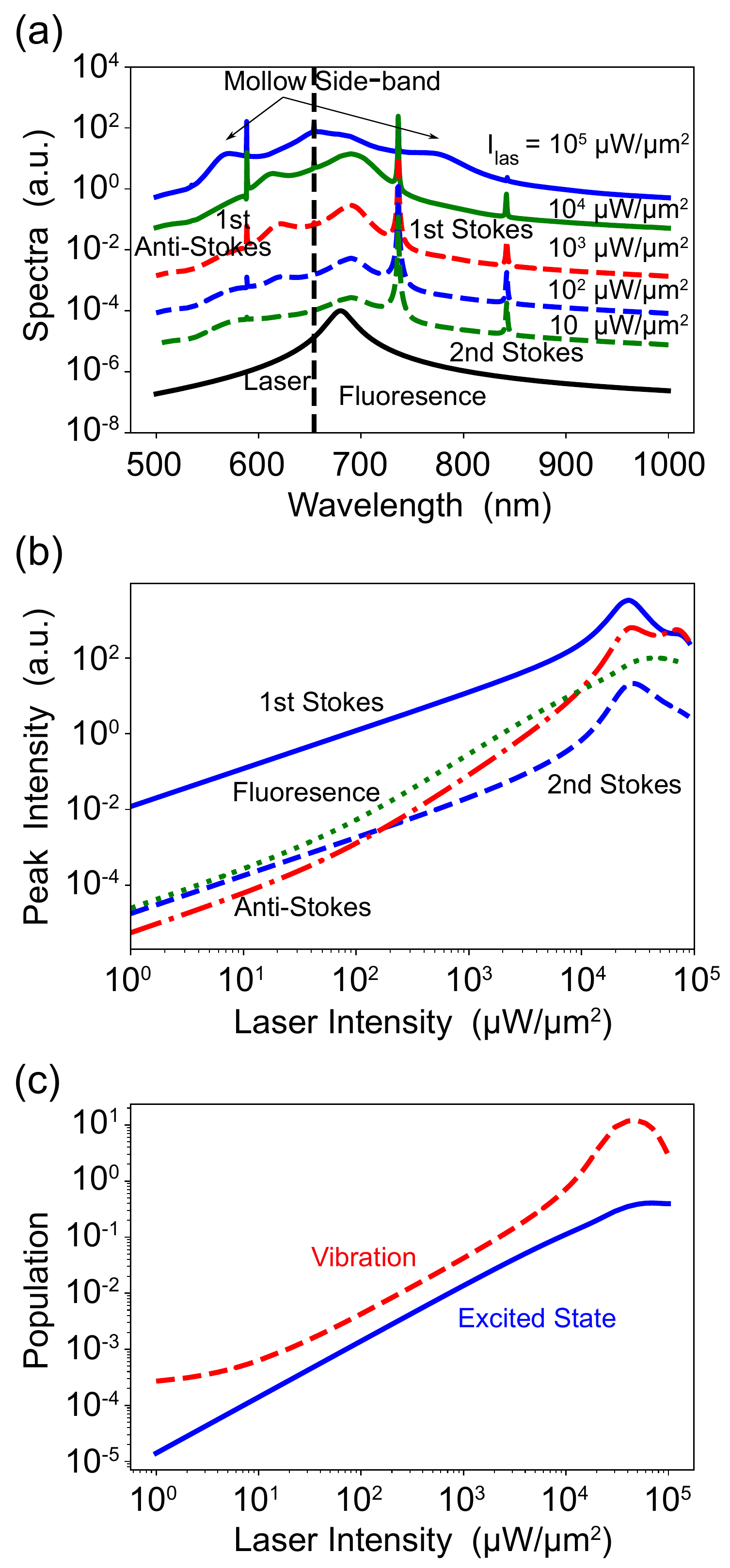}
\par\end{centering}
\caption{ \label{fig:fig6} SERRS response for $\lambda_{\rm las}=654$ nm laser illumination, which is blue-detuned with respect to the Lamb-shifted molecular excitation, and is switched on abruptly at time zero. (a) SERRS spectra for laser intensity $I_{las}=10,10^2,10^3,10^4,10^5 \mu W/\mu m^2$ (lower to upper curves) and fluorescence spectrum in free-space (black curve). (b) The 2nd, 1st order Stokes line (blue dashed and solid line), the anti-Stokes line (red dash-dotted line), and the fluorescence peak (green dotted line) as a function of  $I_{las}$. (c) Electronic excited-state population $\left \langle \hat{\sigma}^\dagger \hat{\sigma}\right \rangle$ and the vibrational population $\left \langle \hat{b}^\dagger_{\nu} \hat{b}_{\nu} \right \rangle$ for the steady-state system as a function of  $I_{las}$. Other parameters are the same as in Fig. \ref{fig:fig4}. }
\end{figure}

\subsection{Blue-detuned Laser Excitation}

We now discuss the system response for $\lambda_{\rm las}=654$ nm laser illumination with increasing intensity $I_{las}$, which is blue-detuned with respect to the Lamb-shifted molecular transition. In Fig. \ref{fig:bd-supp}(a-c) of Appendix \ref{sec:SupRes}, we show that the electronic excited state population $\left \langle \hat{\sigma}^\dagger \hat{\sigma} \right \rangle$, the vibrational population $\left \langle \hat{b}_{\nu}^{\dagger}\hat{b}_{\nu}  \right \rangle$ and the correlation $C(\tau)$ behave similarly as in Fig. \ref{fig:fig4}(a,b) and in Fig. \ref{fig:fig5}(a) for the resonant case, except that: (i)  $\left \langle \hat{\sigma}^\dagger \hat{\sigma} \right \rangle$ oscillates slower for larger $I_{las}$, and the saturated values are smaller; (ii) $\left \langle \hat{b}_{\nu}^{\dagger}\hat{b}_{\nu}  \right \rangle$ shows smaller oscillation but becomes much larger for the largest  $I_{las}$; (iii)  $C(\tau)$ shows only obvious oscillation for the largest $I_{las}$. These differences are partly due to the weaker molecule-near field coupling, which can be attributed itself partially to the lower field enhancement, and partially to the laser-molecular excitation off-resonance. 

Fig. \ref{fig:fig6} (a) shows different spectra in the blue-detuned case, compared to those in the resonant case as shown in Fig. \ref{fig:fig5}(b). More precisely, the anti-Stokes is almost invisible for the lowest laser intensity $I_{las}$, which can be attributed partially to the effects mentioned above and partially to the smaller radiation from the NPoM nanocavity at shorter wavelength (due to the off-resonant condition). For moderate $I_{las}=10^2,10^3,10^4\mu W/\mu m^2$, we also observe one peak at around $656$ nm additionally to the fluorescence peak at $690$ nm. Since this peak evolves to the Mollow side-band, we can attribute this peak to the molecule-laser dressed state, which appears under the off-resonant condition for the weak laser intensity.  For the largest $I_{las}=10^5 \mu W/\mu m^2$ as considered, both Stokes and anti-Stokes lines remain, while the fluorescence peak disappears and a new peak at laser wavelength $\lambda_{las}=654$ nm appears. In addition, the Mollow side peaks depart less from $\lambda_{las}$,  as compared to that in the resonant case for the same laser intensity. 

Fig. \ref{fig:fig6} (b) and (c) show the similar results for the blue-detuned laser illumination as Fig. \ref{fig:fig4} (c) and Fig. \ref{fig:fig5} (c) for the resonant laser illumination except that: (i) the fluorescence intensity scales firstly linearly with increasing $I_{las}$ for $I_{las}\le 10 \mu W/\mu m^2$; (ii) the Stokes and anti-Stokes line increase dramatically for $I_{las}\ge 10^4 \mu W/\mu m^2$, the Stokes line is always stronger than the fluorescence signal, the anti-Stokes line overtakes also the fluorescence signal for larger $I_{las}$; (iii) the vibrational population $\left \langle \hat{b}_{\nu}^\dagger \hat{b}_{\nu} \right \rangle$ increases gradually from the thermal value $n_{\nu}^{th}=2.29\times10^{-4}$ for small laser intensity $I_{las}\le 10 \mu W/\mu m^2 $,  and it increases dramatically over about $10$ for $I_{las}\ge 10^4 \mu W/\mu m^2 $, much larger than that in the resonant case; (iv) the electronic excited-state population $\left \langle \hat{\sigma}^\dagger \hat{\sigma} \right \rangle$ reaches the constant value $0.5$ for $I_{las}\approx 2\times 10^4 \mu W/\mu m^2 $, and it is  smaller than  $\left \langle \hat{b}_{\nu}^\dagger \hat{b}_{\nu} \right \rangle$ for all $I_{las}$. 

The dramatic increase of the vibrational population and the Raman signal at  $I_{las}\ge 10^4 \mu W/\mu m^2$ are due to the stimulated vibrational excitation, which is similar to the stimulated emission in laser operation, and is known as phonon lasing or parametric instability in cavity/molecular optomechanics~\citep{MAspelmeyer,MKSchmidtACSNano}. However, this dramatic increase is limited finally by the suppressed Raman scattering due to the two-level nature of the molecular electronic excitation.

In Fig. \ref{fig:RS-res}(c,d) of Appendix \ref{sec:SupRes}, we have further examined the evolution of the Raman lineshape for the blue-detuned laser illumination, and found that  (i) the Raman shift of the 1st order Stokes lines and anti-Stokes line reduce and increase with same magnitude, respectively, and the reduction of the Raman shift of the 2nd order Stokes line is about twice larger; (ii) the linewidth of the Raman lines reduces for sufficiently larger laser intensity.

\begin{figure}[!htb]
\begin{centering}
\includegraphics[scale=0.53]{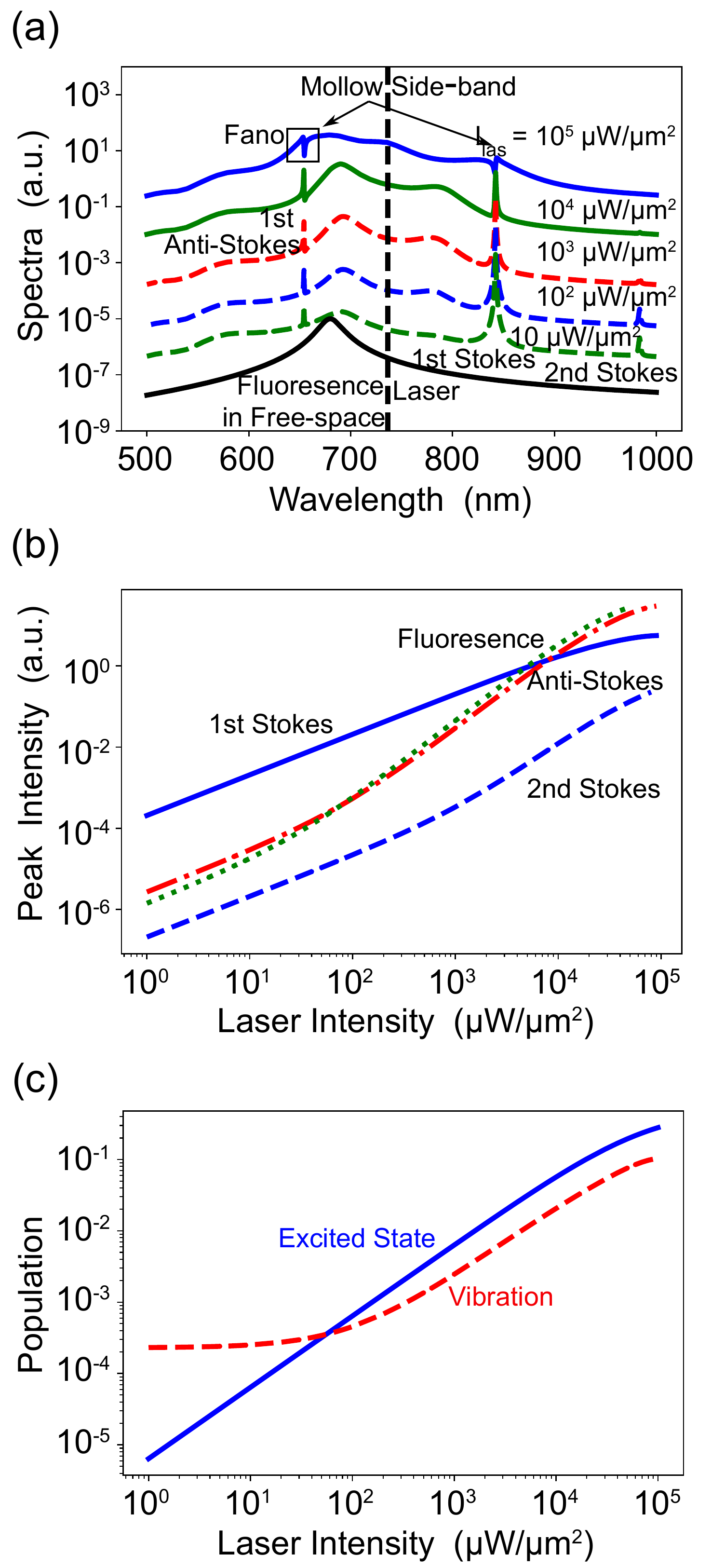}
\par\end{centering}
\caption{ \label{fig:fig7} 
SERRS response for $\lambda_{\rm las}=736$ nm laser illumination, which is red-detuned with respect to the Lamb-shifted molecular excitation. Panel (a), (b), and (c) contain the same information at the corresponding panels in Fig. \ref{fig:fig6}. }
\end{figure}

\subsection{Red-detuned Laser Excitation}

In this section, we complete our analysis by examining in Fig. \ref{fig:fig7} the system response to the  $\lambda_{\rm las}=736$ nm laser illumination, which is red-detuned with respect to the molecular excitation.  In Fig. \ref{fig:bd-supp}(d-f)  of Appendix \ref{sec:SupRes}, we show that the change of the population and the two-time correlation are similar to that of laser illumination blue-detuned with respect to the molecular excitation.

Figure \ref{fig:fig7} (a) shows similar SERRS spectra as those in Fig. \ref{fig:fig6}(a) for the  blue-detuned laser illumination except that: (i) the extra broad peak close to the fluorescence peak for moderate  $I_{las}$  appears at longer wavelength; (ii) for largest $I_{las}=10^5 \mu W/\mu m^2$  there appears two dips at the Stokes and anti-Stokes wavelengths, i.e. Fano features~\citep{TNeumanPRA}. Figure \ref{fig:fig7} (b,c) show a similar change of the spectral intensity and population as in Fig. \ref{fig:fig6} (b,c) for the blue-detuned laser illumination except: (i) the fluorescence and the anti-Stokes line scale linearly with increasing $I_{las}$   for $I_{las} \leq  10^2 \mu W/\mu m^2$, and they become much larger than the Stokes line for $I_{las} \geq  5\times  10^3 \mu W/\mu m^2$; (ii) the vibrational population $\left \langle \hat{b}^\dagger_{\nu} \hat{b}_{\nu} \right \rangle$  remains constant up to a laser intensity $I_{las} \leq  10^2 \mu W/\mu m^2$, about $10$ times larger than that for the blue-detuned illumination,  and it saturates  at $0.3$ for the largest  $I_{las}$ considered instead of increasing dramatically.

In comparison to the cases with resonant and blue-detuned laser illumination, the rate of vibrational pumping is much smaller in this case, and this leads to a relatively larger laser intensity to observe the non-linear anti-Stokes scattering and the vibrational pumping, and to a relatively smaller vibrational population for a given laser intensity. In Fig. \ref{fig:RS-res}(e,f) of Appendix \ref{sec:SupRes}, we have further examined the evolution of the Raman lineshape for the red-detuned laser illumination, and found  that the Raman shift of the Stokes and anti-Stokes lines is increased and decreased, respectively, and the linewidth of the Raman lines starts increasing for laser intensity larger than $2\times 10^2 \mu W/\mu m^2$.

\subsection{Discussions and Conclusions}

In summary, we have developed a theory for surface-enhanced resonant Raman scattering (SERRS) by combining macroscopic quantum electrodynamics theory~\citep{NRivera,SScheel} and electron-vibration interaction in the framework of open quantum system theory, and applied the theory to a realistic system, consisting of a single vertically-standing methylene blue molecule inside a realistic nano-particle on mirror (NPoM) nanocavity. From the calculations, we identified a red-shift of the molecular excitation of about $40$ meV (Lamb shift), and a broadening of the excitation of about $22$ meV (Purcell-enhanced decay rate), and a large molecule-local field interaction for laser intensities currently available in experiments, which leads to Rabi-oscillations, saturated population, and Mollow triplet fluorescence. 

Our theory is formally equivalent to molecular optomechanics in the weak-excitation limit, and thus permits many interesting optomechanical effects in SERRS, such as linear scaling of the vibrational population and super-linear scaling of anti-Stokes signal for moderate laser illumination, super-linear scaling of vibrational population and Raman signal for strong blue-detuned laser illumination among other effects, which makes SERRS attractive for exploration of optomechanical effects in experiments. Since electron-vibration coupling is much larger than the typical optomechanical coupling in molecular optomechanics, the laser intensity to observe these effects is expected to be much smaller in SERRS. In addition, we also observe the 2nd order Stokes line due to the strong electron-vibrational coupling. 

Our theory differs from molecular optomechanics in the aspect that the molecule is effectively a two-level system, but the plasmon has infinite levels as a harmonic oscillator. For strong laser excitation, the molecular excitation is saturated, and this leads to the reduction or even the disappearance of the Raman lines independently of the laser wavelength. Such behavior might provide an explanation to observations in chemical mapping experiments based on SERRS (see Fig. 4d in the supporting information of Ref.~\citep{RZhang}) and in nonlinear resonant Stokes Raman scattering experiments with a MoS$_2$ monolayer inside a NPoM nanocavity (see Fig. 4 of Ref.~\citep{XLiu}). 

In the current study, we have focused on a single vibrational mode and the double resonance of the plasmon and the molecular excitation. Future studies including more vibrational modes and plasmon-molecular excitation detuning~\citep{YZhangNL,RBJaculbia}, might reveal more interesting phenomena. The theory proposed here is valid for a system in the weak coupling regime, as for a single methylene blue molecule in a NPoM nanocavity, but could be further developed for a system in the strong coupling regime, as well as for many methylene blue molecules in a NPoM nanocavity. In the latter case, we expect to observe novel and interesting physics~\citep{TNeumanNP,XLiu} due to the formation of plexcitons, i.e. molecule-plasmon dressed states. 

We have also found that TDDFT calculations for the methylene blue overestimate the transient dipole moment and the Huang-Rhys factors. Thus, the  electronic-vibronic couplings obtained at TDDDFT level are  much larger than those  found in the literature for molecules similar to methylene blue. In future, more precise calculations of the  molecular electronic excited states should be carried out to improve the prediction of the SERRS response.

To conclude, we indicate possible advances of our theory in other aspects if they become significant or necessary for some particular molecules. First, we might consider an-harmonic potential energy surface (PES) to account for the anharmonicity-induced vibrational frequency shift. Second, we might assume different curvatures for the PES of the electronic ground and excited states to account for the different vibrational frequencies for molecules on these states. Finally, we might consider higher excited states of molecules to account for their effect on the SERRS. In any case, the current study establishes a good starting point to investigate the rich physics involved in the nanocavity-enhanced SERRS of a single molecule under strong laser illumination.

\begin{acknowledgments}
We thank Ruben Esteban for the insightful discussion. This work is supported by the project Nr. 12004344, 21902148 from the National Nature Science Foundation of China, joint project Nr. 21961132023 from the NSFC-DPG, project PID2019-107432GB-I00 from the Spanish Ministry of Science and Innovation, and grant IT1526-22 for consolidated groups of the Basque University, through the Department of Education, Research and Universities of the Basque Government. The calculations with Matlab and Gaussian $16$ were performed with the supercomputer at the Henan Supercomputer Center.
\end{acknowledgments}

\section*{Author contributions}
Yuan Zhang devised the theory. Xuan Ming Shen and Yuan Zhang contributed equally to this work. All the authors contribute to the writing of the manuscript.

\newpage
\newpage
\newpage
\newpage

\appendix

\section{Effective Master Equation for Molecule \label{sec:effmaseqn}}

In this appendix, we derive the effective quantum master equation for a single molecule coupled to the quantized electromagnetic field of the metallic nanostructure. According to macroscopic quantum electrodynamics theory~\citep{NRivera,SScheel}, the electromagnetic (plasmonic) field can be described as a continuum via the Hamiltonian 
\begin{equation}
\hat{H}_{f}=\int d\mathbf{r}\int_{0}^{\infty}d\omega_{f}\hbar\omega_{f}\hat{\mathbf{f}}^{\dagger}\left(\mathbf{r},\omega_{f}\right)\cdot\hat{\mathbf{f}}\left(\mathbf{r},\omega_{f}\right)
\end{equation}
with frequency $\omega_{f}$, creation $\hat{\mathbf{f}}^{\dagger}\left(\mathbf{r},\omega_{f}\right)$ and annihilation $\hat{\mathbf{f}}\left(\mathbf{r},\omega_{f}\right)$ noise (bosonic) operators at position $\mathbf{r}$, and the quantized electric field operator reads
\begin{align}
&\hat{\mathbf{E}}\left(\mathbf{r},\omega_{f}\right)=i\sqrt{\frac{\hbar}{\pi\epsilon_{0}}}\frac{\omega_{f}^{2}}{c^{2}}\int d\mathbf{r}'\sqrt{\epsilon^{I}\left(\mathbf{r}',\omega_{f}\right)} \nonumber \\
& \times \overleftrightarrow{G}\left(\mathbf{r},\mathbf{r}';\omega_{f}\right)\cdot\hat{\mathbf{f}}\left(\mathbf{r}',\omega_{f}\right),\label{eq:Eoperator}
\end{align}
with the imaginary part of the dielectric function $\epsilon^{I}\left(\mathbf{r}',\omega\right)$ and the classical dyadic Green's function $\overleftrightarrow{G}\left(\mathbf{r},\mathbf{r}';\omega\right)$. 

We consider the ground and excited electronic state of the molecule, and model them as a two-level system or pseudo-spins, through the Hamiltonian $\hat{H}_{ele}= \hbar \left(\omega_{e}/2\right)\hat{\sigma}^{z},$  with intrinsic transition frequency $\omega_{e}$ and the Pauli operator $\hat{\sigma}^{z}$. In the rotating wave approximation, the molecule interacts with the quantized electric field via the Hamiltonian  $ \hat{H}_{fm}=-\left[\hat{\sigma}^{\dagger}\mathbf{d}_{m}\cdot\hat{\mathbf{E}}\left(\mathbf{r}_{m}\right)+\mathbf{d}_{m}^{*}\cdot\hat{\mathbf{E}}^{\dagger}\left(\mathbf{r}_{m}\right)\hat{\sigma}\right],
$
where $\hat{\mathbf{E}}\left(\mathbf{r}_{m}\right) = \int_{0}^{\infty}d\omega_{f} \hat{\mathbf{E}}\left(\mathbf{r}_{m},\omega_{f}\right)$ is the sum of the field operator given by Eq. \eqref{eq:Eoperator}  for all the frequencies $\omega_{f}$, and $\hat{\sigma},\hat{\sigma}^{\dagger},\mathbf{d}_{m}$ are the lowing and raising operator as well as the transition dipole moment of the molecule.

To reduce the degrees of freedom under consideration, we will treat the electromagnetic field as a reservoir and obtain an effective master equation for the molecule by adiabatically eliminating the reservoir
degree of freedom. To this end, we firstly consider the Heisenberg equation for the operator $\hat{o}$ of the molecule:
\begin{widetext}
\begin{align}
 & \frac{\partial}{\partial t}\hat{o}\left(t\right)=\left[\hat{\sigma}^{\dagger}\left(t\right),\hat{o}\left(t\right)\right]\int_{0}^{\infty}d\omega_{f}\frac{\omega_{f}^{2}}{c^{2}}\int d\mathbf{r}'\sqrt{\frac{\epsilon^{I}\left(\mathbf{r}',\omega_{f}\right)}{\hbar\pi\epsilon_{0}}}\mathbf{d}_{m}\cdot\overleftrightarrow{G}\left(\mathbf{r}_{m},\mathbf{r}';\omega_{f}\right)\cdot\hat{\mathbf{f}}\left(\mathbf{r}',\omega_{f},t\right)\nonumber \\
 & -\mathbf{d}_{m}^{*}\cdot\int_{0}^{\infty}d\omega_{f}\frac{\omega_{f}^{2}}{c^{2}}\int d\mathbf{r}'\sqrt{\frac{\epsilon^{I}\left(\mathbf{r}',\omega_{f}\right)}{\hbar\pi\epsilon_{0}}}\overleftrightarrow{G}^{*}\cdot\left(\mathbf{r},\mathbf{r}';\omega_{f}\right)\hat{\mathbf{f}}^{\dagger}\left(\mathbf{r}',\omega_{f},t\right)\left[\hat{\sigma}\left(t\right),\hat{o}\left(t\right)\right].\label{eq:eq-o}
\end{align}
\end{widetext}
This equation depends on the field operators $\hat{\mathbf{f}}\left(\mathbf{r}',\omega_{f},t\right)$ [and its conjugation $\hat{\mathbf{f}}^{\dagger}\left(\mathbf{r}',\omega_{f},t\right)$], which follows the following Heisenberg equation 
\begin{align}
 & \frac{\partial}{\partial t}\hat{\mathbf{f}}\left(\mathbf{r},\omega_{f},t\right)= -i\omega_{f}\hat{\mathbf{f}}\left(\mathbf{r},\omega_{f},t\right) \nonumber \\
 &+\frac{\omega_{f}^{2}}{c^{2}}\sqrt{\frac{\epsilon^{I}\left(\mathbf{r},\omega_{f}\right)}{\hbar\pi\epsilon_{0}}}\mathbf{d}_{m}^{*}\cdot\overleftrightarrow{G}^{*}\left(\mathbf{r}_{m},\mathbf{r};\omega_{f}\right)\hat{\sigma}\left(t\right),\label{eq:eqn-f}
\end{align}
where we have used the commutation relations $
\left[\hat{\mathbf{f}}\left(\mathbf{r}',\omega_{f}',t\right),\hat{\mathbf{f}}\left(\mathbf{r},\omega_{f},t\right)\right] =0$ and $
\left[\hat{\mathbf{f}}^{\dagger}\left(\mathbf{r}',\omega'_{f},t\right),\hat{\mathbf{f}}\left(\mathbf{r},\omega_{f},t\right)\right]  =-\delta\left(\mathbf{r}-\mathbf{r}'\right)\delta\left(\omega_{f}-\omega_{f}'\right).
$ The formal solution of Eq. (\ref{eq:eqn-f}) is 
\begin{align}
 & \hat{\mathbf{f}}\left(\mathbf{r},\omega_{f},t\right)=\frac{\omega_{f}^{2}}{c^{2}}\sqrt{\frac{\epsilon^{I}\left(\mathbf{r},\omega_{f}\right)}{\hbar\pi\epsilon_{0}}}\mathbf{d}_{m}^{*}\cdot\overleftrightarrow{G}^{*}\left(\mathbf{r}_{m},\mathbf{r};\omega_{f}\right)\nonumber \\
 & \times\int_{0}^{t}dt'e^{-i\omega_{f}\left(t-t'\right)}\hat{\sigma}\left(t'\right).\label{eq:feq-f}
\end{align}
The equation for the conjugate field operator $\hat{\mathbf{f}}^{\dagger}\left(\mathbf{r},\omega_{f},t\right)$ and its formal solution can be achieved by taking the conjugation over Eq. (\ref{eq:eqn-f}) and (\ref{eq:feq-f}). 

At this point, if we insert Eq. (\ref{eq:feq-f}) into Eq. (\ref{eq:eq-o}), we will obtain an integro-differential equation. By solving this equation, we are able to study not only the Markov dynamics in the weak
coupling regime, but also the non-Markov dynamics in the strong coupling regime. Since here we focus on the former regime, we carry out the Born-Markov approximation to the formal solution (\ref{eq:feq-f}).
To do so, we replace $\hat{\sigma}\left(t'\right)$ by $e^{i \omega \left(t-t'\right)}\hat{\sigma}\left(t\right)$ in this expression, and then define a new variable $\tau=t-t'$ to
change the integration over time, and finally change the upper limit of the integration into infinity to achieve the following expression: \begin{align}
 & \hat{\mathbf{f}}\left(\mathbf{r},\omega_{f},t\right)\approx\frac{\omega_{f}^{2}}{c^{2}}\sqrt{\frac{\epsilon^{I}\left(\mathbf{r},\omega_{f}\right)}{\hbar\pi\epsilon_{0}}}\mathbf{d}_{m}^{*}\cdot\overleftrightarrow{G}^{*}\left(\mathbf{r}_{m},\mathbf{r};\omega_{f}\right)\nonumber \\
 & \times\hat{\sigma}\left(t\right)\left(\pi\delta\left(\omega-\omega_{f}\right)+i\mathcal{P}\frac{1}{\omega-\omega_{f}}\right).\label{eq:fBM}
\end{align}
In the last step, we have utilized the relationship 
\begin{equation}
\int_{0}^{\infty}d\tau e^{i\left(\omega-\omega_{f}\right)\tau}=\pi\delta\left(\omega-\omega_{f}\right)+i\mathcal{P}\frac{1}{\omega-\omega_{f}},
\end{equation}
where the symbol $\mathcal{P}$ denotes the principal value. 
Inserting Eq. \eqref{eq:fBM} (and its conjugation) into Eq. \eqref{eq:eq-o}, using the property of the dyadic Green's function
\begin{align}
 & \mathrm{Im}G_{k'k}\left(\mathbf{r}_{1},\mathbf{r}_{2};\omega_{f}\right)  =\frac{\omega_{f}^{2}}{c^{2}}\sum_{j}\int d^{3}\mathbf{r}' \nonumber \\
 &\times \epsilon^{I}\left(\mathbf{r}',\omega_{f}\right)G_{k'j}\left(\mathbf{r}_{1},\mathbf{r}';\omega_{f}\right)G_{kj}^{*}\left(\mathbf{r}_{2},\mathbf{r}';\omega_{f}\right),\label{eq:identity}
\end{align}
and applying the Kramer-Kronig relation 
\begin{align}
 & \mathcal{P}\int d\omega_{f}\frac{1}{\omega_{f}-\omega}\frac{\omega_{f}^{2}}{c^{2}}\mathbf{d}_{m}\cdot\mathrm{Im}\overleftrightarrow{G}\left(\mathbf{r}_{m},\mathbf{r}_{m};\omega_f\right)\cdot\mathbf{d}_{m}^{*}\nonumber \\
 & =\pi\frac{\omega^{2}}{c^{2}}\mathbf{d}_{m}\cdot\mathrm{Re}\overleftrightarrow{G}\left(\mathbf{r}_{m},\mathbf{r}_{m};\omega\right)\cdot\mathbf{d}_{m}^{*},\label{eq:KKrelation}
\end{align}
we obtain the following effective mater equation 
\begin{align}
\frac{\partial}{\partial t}\hat{o}\left(t\right) & =-i\left[\hat{\sigma}^{\dagger}\left(t\right),\hat{o}\left(t\right)\right]\hat{\sigma}\left(t\right)J^{\left(1\right)}\left(\omega_{e}\right)\nonumber \\
 & -iJ^{\left(2\right)}\left(\omega_{e}\right)\hat{\sigma}^{\dagger}\left(\tau\right)\left[\hat{\sigma}\left(t\right),\hat{o}\left(t\right)\right],\label{eq:emq}
\end{align}
with the spectral densities 
\begin{align}
J^{\left(1\right)}\left(\omega_e\right) & =\frac{1}{\hbar\epsilon_{0}}\frac{\omega^{2}}{c^{2}}\mathbf{d}_{m}\cdot\overleftrightarrow{G}\left(\mathbf{r}_{m},\mathbf{r}_{m};\omega_e\right)\cdot\mathbf{d}_{m}^{*},\\
J^{\left(2\right)}\left(\omega_e\right) & =\frac{1}{\hbar\epsilon_{0}}\frac{\omega^{2}}{c^{2}}\mathbf{d}_{m}\cdot\overleftrightarrow{G}^{*}\left(\mathbf{r}_{m},\mathbf{r}_{m};\omega_e\right)\cdot\mathbf{d}_{m}^{*}.
\end{align}
Notice that here we have replaced $\omega$ by $\omega_e$.

In the next step, we consider the equation for the expectation value $\mathrm{tr}\left\{ \hat{o}\left(t\right)\hat{\rho}\right\} =\mathrm{tr}\left\{ \hat{o}\hat{\rho}\left(t\right)\right\} $, which can be computed either with the time-dependent operator $\hat{o}\left(t\right)$ in the Heisenberg picture (left expression) or in the Schr{\"o}dinger picture (right expression). Using this relationship and the cyclic property of the trace ${\rm tr}\{\hat{a}\hat{b}\}={\rm tr}\{\hat{b}\hat{a}\}$ (for any operator $\hat{a},\hat{b}$), we obtain the equation for the reduced density operator 
\begin{equation}
\frac{\partial}{\partial t}\hat{\rho}  =-i\left[\hat{\sigma}\hat{\rho},\hat{\sigma}^{\dagger}\right]J^{\left(1\right)}\left(\omega_{e}\right)\nonumber -iJ^{\left(2\right)}\left(\omega_{e}\right)\left[\hat{\rho}\hat{\sigma}^{\dagger},\hat{\sigma}\right].\label{eq:meq}
\end{equation}
Introducing the new parameters 
\begin{align}
\Omega & =-\frac{1}{\hbar\epsilon_{0}}\frac{\omega_{e}^{2}}{c^{2}}\mathbf{d}_{m}\cdot\mathrm{Re}\overleftrightarrow{G}\left(\mathbf{r}_{m},\mathbf{r}_{m};\omega_{e}\right)\cdot\mathbf{d}_{m}^{*},\\
\Gamma & =\frac{2}{\hbar\epsilon_{0}}\frac{\omega_{e}^{2}}{c^{2}}\mathbf{d}_{m}\cdot\mathrm{Im}\overleftrightarrow{G}\left(\mathbf{r}_{m},\mathbf{r}_{m};\omega_{e}\right)\cdot\mathbf{d}_{m}^{*},
\end{align}
we can rewrite the spectral densities as $J^{\left(1\right)}\left(\omega_{e}\right)=-\Omega+i\Gamma/2,J^{\left(2\right)}\left(\omega_{e}\right)=-\Omega-i\Gamma/2.$
Inserting these expressions into Eq. (\ref{eq:meq}), we achieve the
following effective master equation 
\begin{align}
\frac{\partial}{\partial t}\hat{\rho} & =-i\frac{1}{2}\left(\omega_{e}+\Omega\right)\left[\hat{\sigma}^{z},\hat{\rho}\right]\nonumber \\
 & +\frac{1}{2}\Gamma\left(2\hat{\sigma}\hat{\rho}\hat{\sigma}^{\dagger}-\hat{\sigma}^{\dagger}\hat{\sigma}\hat{\rho}-\hat{\rho}\hat{\sigma}^{\dagger}\hat{\sigma}\right).\label{eq:master-equation}
\end{align}
On the basis of this equation, we arrive at Eq. \eqref{eq:me} in the main text where we include the coupling with the local electric field, the Hamiltonian of the vibrational modes, the electron-vibration coupling as well as the other dissipative processes related to the electronic states and the vibrational modes. 

\section{Far-field Spectrum \label{sec:spe} }

In this appendix, we present the derivation of the far-field radiation from the single molecule in the NPoM nano-cavity. According to Ref.~\citep{MOScully}, the far-field spectrum can be computed with 
\begin{equation}
\frac{dW}{d\Omega}\left(\omega\right)=\frac{c\epsilon_{0}r^{2}}{4\pi^{2}}\mathrm{Re}\int_{0}^{\infty}e^{i\omega\tau}d\tau\mathrm{tr}\left\{ \hat{\mathbf{E}}^{\dagger}\left(\mathbf{r}_d,0\right)\cdot\hat{\mathbf{E}}\left(\mathbf{r}_d,\tau\right)\hat{\rho}\right\} .\label{eq:spectrum}
\end{equation}
In this expression, $r$ is the distance between the molecule and the detector, $\hat{\mathbf{E}}\left(\mathbf{r},\tau\right)=\int_{0}^{\infty}d\omega_{f}\hat{\mathbf{E}}\left(\mathbf{r},\omega_{f},\tau\right)$
is the electric field operator at the detector position $\mathbf{r}_d$, and $\omega$ is the frequency of the spectrum. Inserting Eq. (\ref{eq:fBM}) into Eq. (\ref{eq:Eoperator}), we obtain the following expression
\begin{align}
 & \hat{\mathbf{E}}\left(\mathbf{r},\omega_{f},\tau\right)=i\frac{\hbar}{\pi\epsilon_{0}}\frac{\omega_{f}^{2}}{c^{2}}\mathrm{Im}\overleftrightarrow{G}\left(\mathbf{r},\mathbf{r}_{m};\omega_{f}\right)\cdot\mathbf{d}_{m}^{*}\nonumber \\
 & \times\hat{\sigma}\left(\tau\right)\left(\pi\delta\left(\omega_{f}-\omega \right)+i\mathcal{P}\frac{1}{\omega -\omega_{f}}\right).
\end{align}Here, we have utilized the relation (\ref{eq:identity}), and replaced $\omega$ by $\omega_{f}$. Using the above expression and Eq. (\ref{eq:KKrelation}), we obtain the following expression for the electric field operator 
\begin{equation}
\hat{\mathbf{E}}\left(\mathbf{r},\tau\right)=\frac{\hbar}{\epsilon_{0}}\frac{\omega^{2}}{c^{2}}\overleftrightarrow{G}\left(\mathbf{r},\mathbf{r}_{m};\omega\right)\cdot\mathbf{d}_{m}^{*}\hat{\sigma}\left(\tau\right).
\end{equation}Applying the conjugation to the above equation, we can obtain the expression for the conjugated field operators $\hat{\mathbf{E}}^{\dagger}\left(\mathbf{r},\tau\right)$. Inserting these results into Eq. (\ref{eq:spectrum}), we can rewrite the spectrum as 
\begin{equation}
\frac{dW}{d\Omega}\left(\omega\right)\approx K(\omega) \mathrm{Re}\int_{0}^{\infty}d\tau e^{i\omega\tau}\mathrm{tr}\left\{ \hat{\sigma}^{\dagger}\left(0\right)\hat{\sigma}\left(\tau\right)\hat{\rho}\right\} .\label{eq:spectrum-bm}
\end{equation}with the propagation factor
\begin{equation}
K(\omega)=\frac{\hbar^{2}cr^{2}}{4\pi^{2}\epsilon_{0}}\left|\frac{\omega^{2}}{c^{2}}\overleftrightarrow{G}\left(\mathbf{r},\mathbf{r}_{m};\omega\right)\cdot\mathbf{d}^{*}_{m}\right|^{2}.
\end{equation}

To compute the spectrum with Eq. (\ref{eq:spectrum-bm}), we need to evaluate the two-time correlations $\mathrm{tr}\left\{ \hat{\sigma}^{\dagger}\left(0\right)\hat{\sigma}\left(\tau\right)\hat{\rho}\right\} $, where $\tau$ refers the difference of time with respect to the steady-state labeled as ``$0$''. To compute these correlations, we consider a pure quantum system. In this case, we can introduce the time-dependent propagation operator $\hat{U}\left(\tau\right)$ to reformulate the correlations as
\begin{align}
 & \mathrm{tr}\left\{ \hat{\sigma}^{\dagger}\left(0\right)\hat{\sigma}\left(\tau\right)\hat{\rho}\right\} =\mathrm{tr}\left\{ \hat{\sigma}^{\dagger}\hat{U}^{\dagger}\left(\tau\right)\hat{\sigma}\hat{U}\left(\tau\right)\hat{\rho}\right\} \nonumber \\
 & =\mathrm{tr}\left\{ \hat{\sigma}\hat{U}\left(\tau\right)\hat{\rho}\hat{\sigma}^{\dagger}\hat{U}^{\dagger}\left(\tau\right)\right\} =\mathrm{tr}\left\{ \hat{\sigma}\hat{\varrho}\left(\tau\right)\right\},
\end{align}where we have defined the operator $\hat{\varrho}\left(\tau\right)=\hat{U}\left(\tau\right)\hat{\rho}\hat{\sigma}^{\dagger}\hat{U}^{\dagger}\left(\tau\right)$. In essence, we have transformed the expression in the Heisenberg picture to that in the Schr{\"o}dinger picture. To deal with the quantum system in the presence of loss, we should replace $\hat{U}\left(\tau\right)...\hat{U}^{\dagger}\left(\tau\right)$ with the time-dependent propagation superoperator $\hat{\mathcal{U}}\left(\tau\right)$, which indicates the formal solution of the master equation, such as  Eq. (\ref{eq:master-equation}). Finally, we can compute the spectrum as 
\begin{equation}
\frac{dW}{d\Omega}\left(\omega\right)\approx K(\omega) \mathrm{Re}\int_{0}^{\infty}d\tau e^{i\omega\tau}\mathrm{tr}\left\{ \hat{\sigma}\hat{\varrho}\left(\tau\right)\right\} ,\label{eq:spec-Schordinger}
\end{equation}
where $\hat{\varrho}\left(\tau\right)$ satisfies the same master equation as $\hat{\rho}$, however with the initial condition $\hat{\varrho}\left(0\right)=\hat{\rho}\hat{\sigma}^{\dagger}$.

\section{Supplemental Results \label{sec:SupRes}}

In this appendix, we present and discuss extra information to supplement the results in the main text. 

\begin{figure*}[!htb]
\begin{centering}
\includegraphics[scale=0.35]{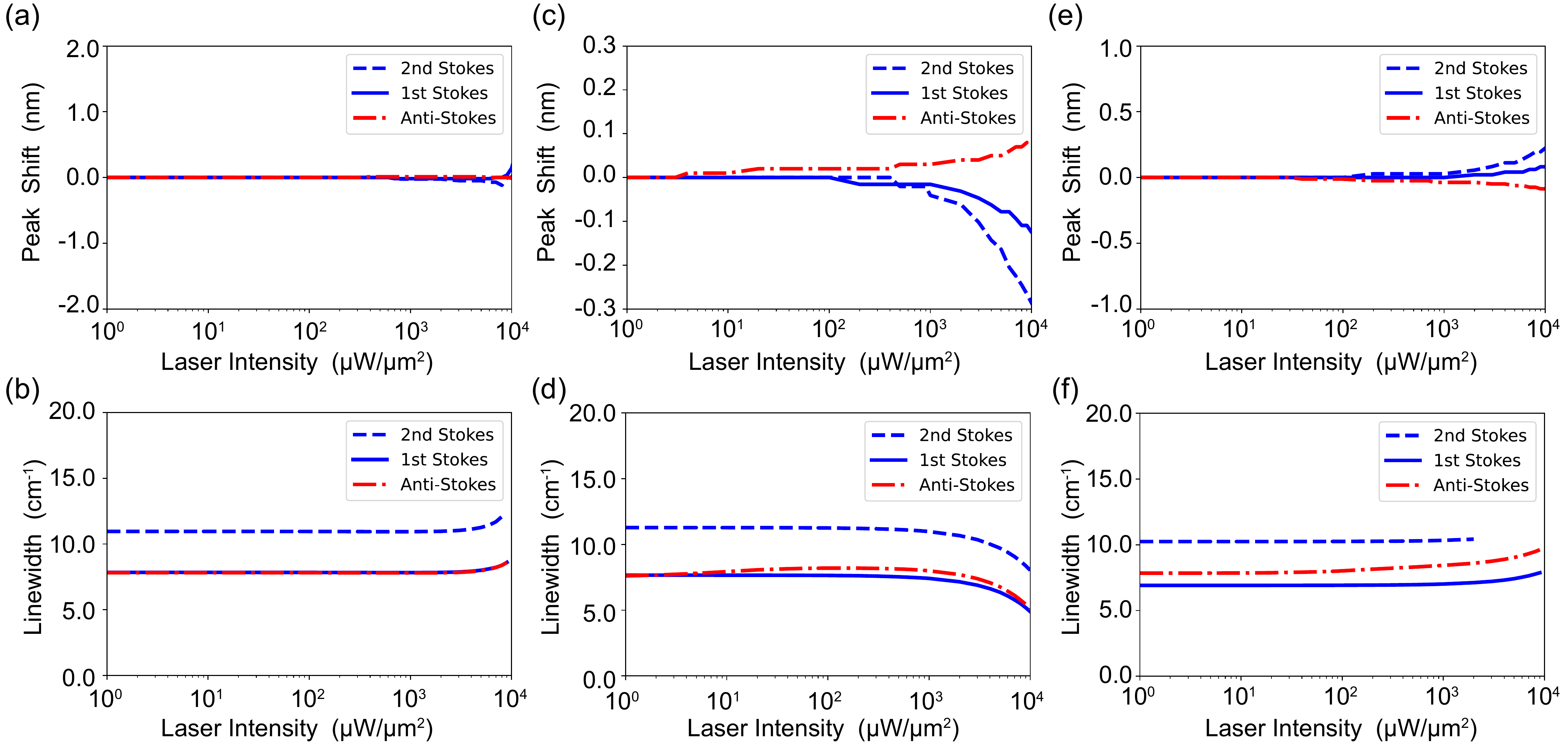}
\par\end{centering}
\caption{ \label{fig:RS-res} Evolution of  the change of the Raman shift (a,c,e) and the Raman linewidth (b,d,f)  for laser illumination with increasing laser intensity $I_{las}$ and wavelength  $\lambda_{las}=693$ nm (a,b),  $\lambda_{las}=654$ nm (c,d),  $\lambda_{las}=736$ nm (e,f), which are  resonant, blue- and red-detuned to the Lamb-shifted molecular excitation at $693$ nm, respectively. Other parameters are the same as in Fig. \ref{fig:fig4}.}
\end{figure*}

Figure \ref{fig:RS-res}(a,b) show the change of Raman lineshape for laser illumination of increasing intensity $I_{las}$, which is resonant to the molecular excitation. We see that the frequency of the Raman peaks do not change while the linewidth gets broadened for larger $I_{las}$. In addition, we find that  the 2nd order Stokes line is about twice broader than the 1st order Stokes line.


Figure \ref{fig:RS-res}(c,d) show the change of Raman lineshape for laser illumination blue-detuned with respect to the molecular excitation. Figure \ref{fig:RS-res}(c)  shows that the Raman shift of the Stokes lines is reduced, and the shift reduction is twice larger for the 2nd order Stokes line, while the Raman shift of the anti-Stokes line is increased and the magnitude of the shift change is comparable to that of the 1st order Stokes line.  Figure \ref{fig:RS-res}(d) shows that the linewidth of the Raman lines starts reducing for laser intensity larger than $10^3 \mu W/\mu m^2$.


Figure \ref{fig:RS-res}(e,f) show the change of Raman lineshape for laser illumination red-detuned with respect to the molecular excitation. Figure \ref{fig:RS-res}(e)  shows that the Raman shift of the Stokes and anti-Stokes lines is increased and decreased, respectively.  Figure \ref{fig:RS-res}(f) shows that the linewidth of the Raman lines starts increasing for laser intensity larger than $2\times 10^2 \mu W/\mu m^2$.

\begin{figure*}
\begin{centering}
\includegraphics[scale=0.4]{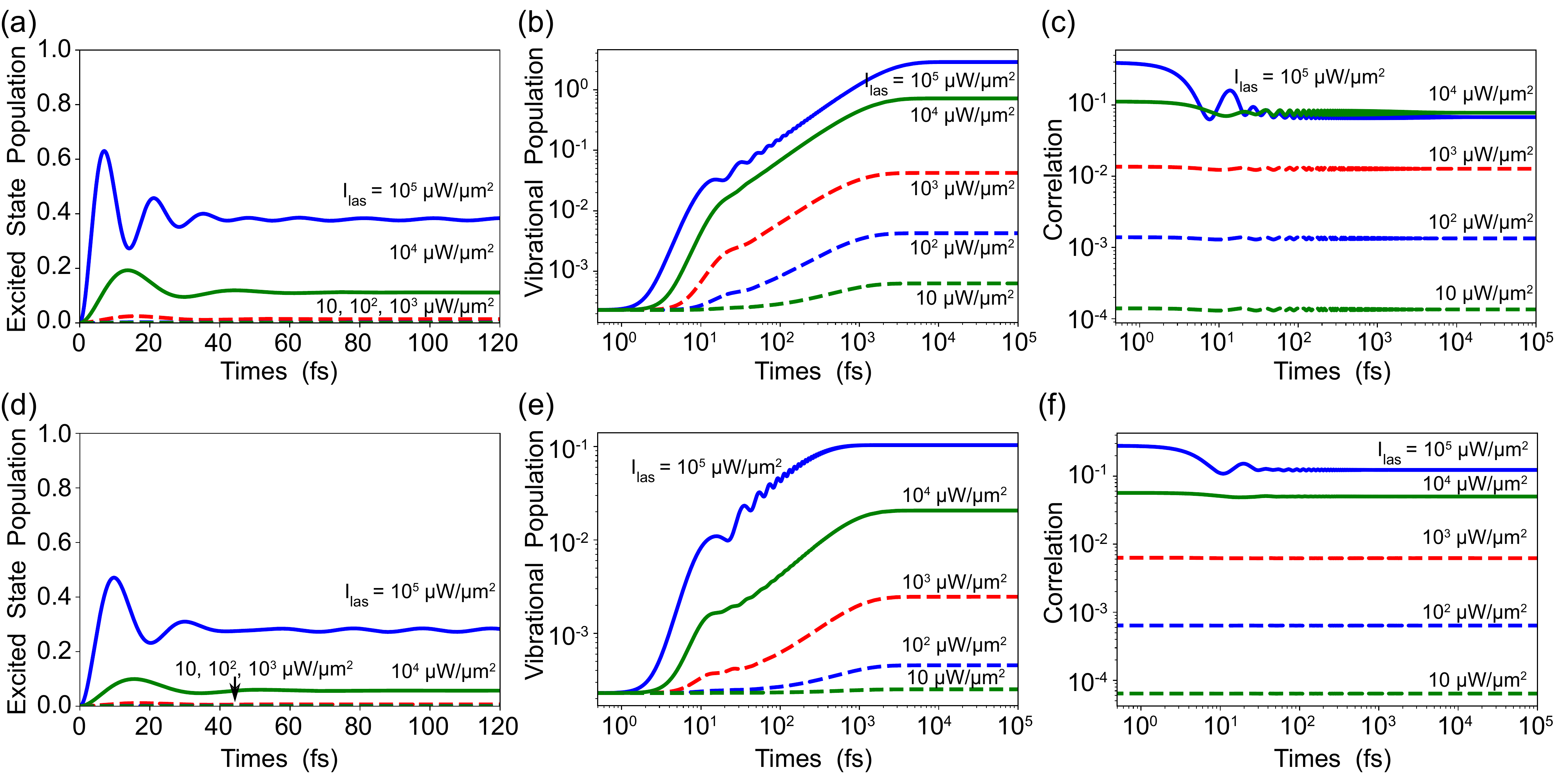}
\par\end{centering}
\caption{ \label{fig:bd-supp} Electronic excited-state population (a,d), vibrational population (b,e), and two-time correlation (c,f) for the $654$ nm (a,b,c) and $736$ nm (d,e,f) laser illumination with increasing intensity $I_{las}$, which are  blue- and red-detuned to the Lamb-shifted molecular excitation, respectively. Other parameters are the same as in Fig. \ref{fig:fig4} in the main text.}
\end{figure*}

Figure \ref{fig:bd-supp}(a,b,c) show that the dynamics of the electronic excited state population (a), vibrational population (b), and two-time correlation (c) for $654$ nm laser illumination, blue-detuned to the molecular excitation, are similar to those for $693$ nm laser excitation resonant to the molecular excitation, as shown in Fig. \ref{fig:fig4}(a,b) and in Fig. \ref{fig:fig5}(a), except: (i) the excited-state population is weaker and its oscillation is slower; (ii) the vibrational population is much larger for a given laser intensity.


Figure  \ref{fig:bd-supp}(d,e,f) show that the dynamics of the electronic excited state population (d), vibrational population (e), and two-time correlation (f) for $736$ nm laser illumination, red-detuned to the molecular excitation, are similar to those for $654$ nm laser excitation, blue-detuned with respect to the molecular excitation, except that the vibrational population is relatively weak for a given laser intensity.

\end{document}